\pdfoutput=1
\documentclass[twocolumn,amsmath,amssymb,prd,superscriptaddress,nofootinbib,floatfix,showpacs,preprintnumbers]{revtex4-1}
\usepackage{graphicx}
\usepackage{color}
\usepackage{hyperref}
\usepackage{dcolumn}
\usepackage{subfigure}
\usepackage{xspace}

\newcommand{\onefig}{1.1\linewidth}
\newcommand{\twofigs}{0.4\linewidth}

\newcommand{\pT}{\ensuremath{T_0}}

\newcommand{\sigx}{\ensuremath{\sigma_x}}
\newcommand{\sigy}{\ensuremath{\sigma_y}}

\newcommand{\einh}[1]{\ensuremath{\,\text{#1}}}
\newcommand{\MeV}{\einh{MeV}}

\newcommand{\ua}{\ensuremath{U(1)_A}}
\newcommand{\Phibar}{\ensuremath{\bar{\Phi}}}
\newcommand{\LPQM}{\ensuremath{\mathcal{L}_{\textrm{PQM}}}\xspace}
\newcommand{\dslash}{\ensuremath{\partial\hspace{-1.2ex} /}}
\newcommand{\Dslash}{\ensuremath{D\hspace{-1.5ex} /}}
\newcommand{\Tr}{\ensuremath{\operatorname{Tr}}}

\def\Eq#1{Eq.~(\ref{#1})}
\def\Fig#1{Fig.~\ref{#1}}
\def\Tab#1{Tab.~\ref{#1}}

\graphicspath{
{./}
{../figures/}
}

\begin{document}
\title{QCD critical region and higher moments for three flavor models}
\author{B.-J. Schaefer}
\email{bernd-jochen.schaefer@uni-graz.at}
\affiliation{Institut f\"{u}r Theoretische Physik, Justus-Liebig-Universit\"{a}t Giessen, D-35392 Giessen, Germany}
\affiliation{Institut f\"{u}r Physik, Karl-Franzens-Universit\"{a}t Graz,
 A-8010 Graz, Austria} \author{M. Wagner}
\email{mwagner@physik.uni-bielefeld.de}
\affiliation{Fakult\"at f\"ur Physik, Universit\"at Bielefeld, D-33615
  Bielefeld, Germany}

\date{\today}
\preprint{BI-TP 2011/45}

\begin{abstract}
  One of the distinctive features of the QCD phase diagram is the
  possible emergence of a critical endpoint. The critical region
  around the critical point and the path dependency of the critical
  exponents are investigated within effective chiral $(2+1)$-flavor
  models with and without Polyakov loops. Results obtained in no-sea
  mean-field approximations where a divergent vacuum part in the
  fermion-loop contribution is neglected are compared to the
  renormalized ones. Furthermore, the modifications caused by the
  backreaction of the matter fluctuations on the pure Yang-Mills
  system are discussed. Higher-order, non-Gaussian moments of
  event-by-event distributions of various particle multiplicities are
  enhanced near the critical point and could serve as a probe to
  determine its location in the phase diagram. By means of a novel
  derivative technique higher-order generalized quark-number
  susceptibilities are calculated and their sign structure in the
  phase diagram is analyzed.
\end{abstract}

\maketitle

\section{Introduction}

Two important properties of the QCD vacuum are spontaneous chiral
symmetry breaking and color confinement. It is expected that chiral
symmetry can be restored and that a confinement/deconfinement transition
occurs at high temperature and/or baryon densities
\cite{Fukushima:2010bq, *Fukushima:2011jc}.  A deeper understanding
of how these phase transitions manifest themselves both theoretically and
experimentally is of utmost importance in mapping the phase diagram
\cite{Leupold:2011zz}.

A crossover transition at vanishing density seems to be well
established by recent QCD lattice simulations at almost physical quark
masses.  With a better continuum extrapolation and improved actions
the latest value of the chiral critical temperature $T_\chi \approx
154\pm9$ MeV by the HotQCD Collaboration~\cite{Bazavov:2011nk} is in agreement with $T_\chi \approx 147-157 \MeV$ of
the Wuppertal-Budapest group~\cite{Borsanyi:2010bp, *Aoki:2006br,
  *Aoki:2009sc}. However, the extrapolations of lattice simulations
towards finite chemical potential are much under debate.
Most theoretical model studies evidence a genuine second-order
critical endpoint (CEP) where the first-order chiral phase transition
line at large chemical potential terminates~\cite{Fukushima:2010pp,
  Aoki:2006we, *Gupta:2011wh, *Son:2004iv}. Many properties of the
endpoint such as, e.g., its exact location remain unknown.  QCD
lattice simulations cannot directly access the relevant region in the
phase diagram but extensions towards finite chemical potential provide
some limitations and can rule out the existence of a CEP for small
$\mu/T$ ratios \cite{Philipsen:2011zx}.
This observation agrees with recent first-principle QCD studies
performed with functional renormalization group techniques and
Dyson-Schwinger equations~\cite{Herbst:2010rf, Braun:2009gm,
  Fischer:2011mz}.  Moreover, the predicted chiral first-order
transition at small temperatures is model dependent.  For example, the
inclusion of the 't Hooft anomaly in chiral models may change the
first-order transition to a smooth crossover~\cite{Hatsuda:2006ps,
  *Chen:2009gv, *Yamamoto:2007ah}.

This underlines the importance of the planned facilities such as the
Compressed Baryonic Matter (CBM) experiment at the Facility for
Antiproton and Ion Research (FAIR) and the Nuclotron-Based Ion
Collider Facility (NICA) at the Joint Institute for Nuclear Research
(JINR) and of the recently started "beam energy scan" at the
Relativistic Heavy Ion Collider (RHIC, Brookhaven National Laboratory)
\cite{Aggarwal:2010wy, *Adams:2005dq} which are dedicated to search
for a critical point.  For an experimental overview see
e.g.~\cite{Mohanty:2009vb}.  The knowledge of the characteristic
signatures of the conjectured CEP is inevitable for the experiments
\cite{Koch:2008ia, Luo:2011rg}.  Possible experimental signatures
which are based on the singular behavior of thermodynamic functions
near a critical point were already suggested a decade ago (see,
e.g.,~\cite{Stephanov:1999zu, *Stephanov:1998dy}). They are related to
temperature and chemical potential fluctuations in event-by-event
fluctuations of various particle multiplicities \cite{Jeon:2000wg,
  *Jeon:2003gk}. By scanning the center of mass energy and thus the
baryochemical potential an increase and then a decrease in the number
fluctuations of, e.g., pions and protons should be seen as one crosses
the critical point. The nonmonotonic behavior in the number
fluctuations in the vicinity of a critical point might serve as a
probe to determine its location in the phase diagram assuming that the
signals are not washed out by the expansion of the colliding system
\cite{Ejiri:2005wq}.

In a realistic heavy-ion collision the correlation length is cut off by
critical slowing down and finite volume effects.
In addition, the system has only a finite time to build up
correlations.  Estimates of the largest correlation length as the
collision cools past the critical point are in the range of a few
$(2-3)$ fm and only a factor 3 larger than the natural scale
$(0.5-1)$ fm far away from the critical point~\cite{Berdnikov:1999ph}.
Hence, if the correlation length increases only by a few percent near
a critical point more sensible quantities are needed for the analysis
of freeze-out and critical conditions in heavy-ion
collisions~\cite{Stephanov:2008qz, Athanasiou:2010kw}. Possible
candidates are higher-order cumulants or ratios of higher-order
generalized susceptibilities which depend on higher powers of the
correlation length $\xi$. An example is the fourth-order
susceptibility $\chi_4$ which scales like $\xi^7$ near the critical
point; hence, an enhancement by a factor 2 is expected for this
quantity. The sign of the fourth moment should also change as the CEP
is approached from the crossover side as pointed out in
Ref.~\cite{Gavai:2010zn, *Stephanov:2011pb}.

In this work we investigate the sign structure not only of the fourth
moment (kurtosis) but also of higher-order moments in the $(T,\mu)$
phase diagram. We discuss the generic behavior of these higher moments
at nonvanishing chemical potential within ($2+1$)-flavor quark-meson
(QM) and Polyakov-quark-meson (PQM) models which we briefly
recapitulate in Sec.~\ref{sec:pqm}. The no-sea mean-field findings are
compared to the results obtained in the renormalized models.  As
mentioned in \cite{Hatta2003, Schaefer:2006ds} the critical region
around the critical point in the phase diagram is not pointlike but
has a richer structure. In Sec.~\ref{sec:phasestructure} we
investigate the size of the critical region and the path dependency of
the critical exponents towards the critical point.  The influence of
the backreaction of matter fluctuations on the pure Yang-Mills system
on the critical region is analyzed. In Sec.~\ref{sec:moments}
different ratios of moments of the baryon number fluctuations to high
orders are calculated by means of a novel derivative technique
\cite{Wagner:2009pm}. We conclude and point out some experimental
consequences from these findings in Sec.~\ref{sec:summary}.

\section{Three-flavor models}
\label{sec:pqm}

Chiral symmetry and its spontaneous breaking for three flavor is very
well described by renormalizable quark-meson type models which serve
as an effective realization of low-energy strongly-interacting matter
\cite{Roder:2003uz, *Herpay:2005yr, *Lenaghan:2000ey,
  *Scavenius:2000qd, *Schaefer1999, *Meyer-Ortmanns:1994nt}.  These
models can be augmented with the Polyakov loop (see, e.g., \cite{
  Fukushima:2003fw, *Meisinger:2002kg,
  *Megias:2004hj,*Megias:2006bn,Ratti:2005jh}), yielding the PQM
models which mimic in addition to the chiral symmetry breaking certain
aspects of a statistical confinement.

The Lagrangian of the three-flavor ($N_f=3$) PQM model
\cite{Schaefer:2009ui, *Mao:2009aq, *Gupta:2009fg, *Chatterjee:2011jd} is
assembled from three parts
\begin{equation}
  \label{eq:lpqm}
  \LPQM = \bar{q}\left(i \Dslash + h \phi_5 + \gamma_0 \mu_f \right) q + \mathcal{L}_m
  -\mathcal{U} (\Phi, \Phibar)\ ,
\end{equation}
where the first term describes the interaction between the $N_c=3$
color quark fields $q = (u,d,s)$ and the mesons $\phi_5 = T_a
(\sigma_a + i\gamma_5 \pi_a)$ with a flavor-blind Yukawa coupling
$h$. By $T_a$ the nine generators of the $U(3)$ symmetry are denoted
and $\sigma_a$ ($\pi_a$) labels the (pseudo)scalar meson nonets. For
each flavor $f \in \{u,d,s \}$ an independent quark chemical potential
$\mu_f$ is introduced but in this paper we will consider symmetric
quark matter with one uniform quark chemical potential $\mu \equiv
\mu_B/3$.  The quarks are coupled to a temporal, spatially constant
background gauge field $A_0$, which is represented in terms of the
Polyakov loops, via a covariant derivative $\Dslash = \dslash - i
\gamma_0 A_0$. A gauge coupling has been absorbed in the gauge fields.

The second part in the Lagrangian, $\mathcal{L}_m$ represents the
purely mesonic contribution with the field \mbox{$\phi = T_a (\sigma_a +
i\pi_a)$} and reads
\begin{eqnarray}
  \label{eq:mesonL}
  \mathcal{L}_m &=& \Tr \left( \partial_\mu \phi^\dagger \partial^\mu
    \phi \right)
  - m^2 \Tr ( \phi^\dagger \phi) -\lambda_1 \left[\Tr (\phi^\dagger
    \phi)\right]^2 \nonumber \\
  &&  - \lambda_2 \Tr\left(\phi^\dagger \phi\right)^2
  +c   \left(\det (\phi) + \det (\phi^\dagger) \right)\nonumber \\
  && +\Tr\left[H(\phi+\phi^\dagger)\right]\ .
\end{eqnarray}

The last term in \Eq{eq:mesonL} breaks chiral symmetry explicitly
where $H = T_a h_a$ has in general nine external parameters $h_a$
whereof only two are nonvanishing for a $(2+1)$-flavor symmetry
breaking pattern.  The $\ua$ symmetry is broken explicitly by the 't
Hooft determinant with a constant strength $c$.

The last part of \Eq{eq:lpqm} represents the effective gluon field
potential in terms of the Polyakov-loop variables $\Phi$ and $\Phibar$
\cite{Pisarski:2000eq, *Dumitru:2001xa} which approximate the
dynamical glue sector of QCD.  For vanishing chemical potential the
bulk thermodynamics of lattice Yang-Mills is reproduced up to twice
the deconfinement phase transition temperature~\cite{Schaefer:2008ax,
  Schaefer:2009st,*Wambach:2009ee, Schaefer:2009ui}. Several explicit
choices of the Polyakov-loop potential are known in the literature
\cite{Ratti:2005jh, Roessner:2006xn, Fukushima:2008wg}, see
\cite{Schaefer:2009ui} for a comparison. In this work we employ the
logarithmic version \cite{Roessner:2006xn}
\begin{eqnarray}
  \label{eq:ulog}
  \frac{\mathcal{U}_{\text{log}}}{T^{4}} &=& -\frac{a(T)}{2} \Phibar \Phi\\
  &&+ b(T) \ln \left[1-6 \Phibar\Phi + 4\left(\Phi^{3}+\Phibar^{3}\right)
    - 3 \left(\Phibar \Phi\right)^{2}\right]\ ,\nonumber
\end{eqnarray}
with temperature-dependent coefficients
\begin{equation*}
  a(T) =  a_0 + a_1 \left(\frac{T_0}{T}\right) + a_2
  \left(\frac{T_0}{T}\right)^2\ , \
  b(T) = b_3 \left(\frac{T_0}{T}\right)^3
\end{equation*}
and the parameters $a_0 = 3.51$, $a_1 = -2.47$, $a_2 = 15.2$ and $b_3
= -1.75$ \cite{Schaefer:2009ui}. Originally, the parameter $T_0$ is
fitted to the critical temperature $T_0=270$ MeV of the first-order
deconfinement transition of the quenched $SU(3)$ lattice gauge theory.

However, in full dynamical QCD the effective Polyakov-loop potential
has to be replaced by the corresponding QCD Yang-Mills (YM) part.
Within a functional renormalization group approach the critical
temperature of the quenched YM system could be determined recently
\cite{Braun:2007bx}. Because of the matter fluctuations the glue
propagation is further dressed (see also \cite{Braun:2009gm}). A first
phenomenological hard thermal loop estimate for this flavor and
chemical potential dependence of the critical temperature has been
given in \cite{Schaefer:2007pw} and within perturbation theory (see
\cite{Braun:2005uj, *Braun:2006jd}).  These considerations demonstrate
that low-energy models such as the PQM model can be understood as a
specific approximation of QCD and can be systematically enhanced
towards full dynamical QCD.

To be more specific: the matter backreaction to the gluon sector
leads to $N_f$ and $\mu$ modifications in $T_0$, i.e., $T_0 \to T_0
(N_f, \mu)$ as introduced in \cite{Schaefer:2007pw} (see also
\cite{Braun:2005uj, *Braun:2006jd, Schaefer:2009ui, Herbst:2010rf,
  Schaefer:2011pn, *Pawlowski:2010ht} for more details).

 Explicitly, we employ
\begin{equation}
  \label{eq:t0mu}
  \pT(N_f, \mu) = T_{\tau} e^{-1/(\alpha_0 b(N_f, \mu))}
\end{equation}
where $T_\tau = 1.77$ GeV denotes the $\tau$ scale and
$\alpha_0=\alpha(\Lambda)$ the gauge coupling at some UV scale
$\Lambda$. The $\mu$-dependent running coupling reads
\begin{equation}
 b(N_f, \mu) = b(N_f) - b_{\mu}\frac{\mu^2}{ T_{\tau}^2}\ ,
\end{equation}
with the factor $b_{\mu}\simeq \frac{16}{\pi}N_f$.

Finally, the grand potential in mean-field approximation (MFA) is the
sum of three terms consisting of the meson, $U$, the fermion-loop
contribution $\Omega_{\bar{q}{q}}$, evaluated in the presence of the
Polyakov loop and the pure gauge field contribution, the Polyakov-loop
potential $\mathcal{U}$:
\begin{equation}
  \label{eq:grandpot}
  \Omega = U \left( \sigx,\sigy\right) +
  \Omega_{\bar{q}{q}} \left({\sigx},{\sigy},
    \Phi,\Phibar \right) +
  \mathcal{U}\left(\Phi,\Phibar\right) \ .
\end{equation}
$\Phi$ and $\Phibar$ denote the real Polyakov-loop expectation values
and ${\sigx}$ and ${\sigy}$ are the light and strange chiral
condensates in the rotated nonstrange-strange basis (see
\cite{Schaefer:2008hk} for further details). The temperature and
density behavior of the four order parameters is obtained by
minimization of the grand potential \Eq{eq:grandpot}.  Explicitly, the
mesonic part reads
\begin{multline}\label{eq:umeson}
  U(\sigx,\sigy) = \frac{m^{2}}{2}\left(\sigma_{x}^{2} +
  \sigma_{y}^{2}\right) -h_{x} \sigma_{x} -h_{y} \sigma_{y}
 - \frac{c}{2 \sqrt{2}} \sigma_{x}^2 \sigma_{y}
\\
  + \frac{\lambda_{1}}{2} \sigma_{x}^{2} \sigma_{y}^{2}+
  \frac{1}{8}\left(2 \lambda_{1} +
    \lambda_{2}\right)\sigma_{x}^{4}+\frac{1}{8}\left(2 \lambda_{1} +
    2\lambda_{2}\right) \sigma_{y}^{4}\ ,
\end{multline}
with six parameters $m^{2}, c, \lambda_{1}, \lambda_{2}, h_{x}, h_{y}$
and the quark-antiquark contribution in the presence of the Polyakov
loop accordingly
\begin{widetext}
  \begin{multline} \label{eq:Omegaqq} \Omega_{\bar qq}(
   \sigx , \sigy, \Phi,\Phibar ) =-2
    \sum_{f=u,d,s} \int\frac{d^3p}{(2\pi)^3} \left\{ N_c E_{q,f}+  T \ln \left[1 + 3
        (\Phi + \bar \Phi e^{-(E_{q,f}-\mu_f)/T})e^{-(E_{q,f}-\mu_f/T} +
        e^{-3(E_{q,f}-\mu_f)/T}\right]\right.\\
    + \left. T\ln \left[1 + 3 (\bar \Phi + \Phi
        e^{-(E_{q,f}+\mu_f)/T})e^{-(E_{q,f}+\mu_f)/T} +
        e^{-3(E_{q,f}+\mu_f)/T}\right] \right\}\
  \end{multline}
\end{widetext}
with the flavor-dependent single-particle energies
\begin{equation}
  E_{q,f}= \sqrt{p^2 + m_f^2}
\end{equation}
and quark masses $m_l = h \sigx /2$ and $m_s = h \sigy /\sqrt{2}$.
The first term in quark/antiquark-loop contribution $\Omega_{\bar
  qq}$, \Eq{eq:Omegaqq}, represents the ultraviolet divergent vacuum
contribution. It is neglected in the no-sea MFA but can be
renormalized with, e.g., the dimensional regularization scheme
yielding
\begin{equation}
\Omega^\text{vac}_{\bar qq}(\sigx , \sigy) = - \frac{N_c}{8\pi^2}
\sum_{f=u,d,s} m_f^4 \log \left( \frac{m_f}{\Lambda}\right)
\end{equation}
where the regularization scale $\Lambda$ has been introduced.  Note
that the $\Lambda$ dependence is completely absorbed by the mesonic
potential $U$.  Hence all thermodynamic quantities are finally
independent of the choice of the regularization scale at the
mean-field level.  With this term fermion fluctuations are taken into
account whereas meson fluctuations are still ignored.  A careful
investigation of the influence of this vacuum term in two-flavor model
studies can be found in~\cite{Andersen:2011pr, Skokov:2010sf,
  Gupta:2011ez} and recently for a $(2+1)$-flavor PQM model in
\cite{Chatterjee:2011jd}. To disentangle physical effects driven by
fluctuations we will compare no-sea mean-field results with the
renormalized ones where this term has been taken into account.

The parameters of the model, \Eq{eq:lpqm}, are chosen to reproduce
experimentally known quantities like the pion and kaon decay constants
and the scalar and pseudoscalar meson masses.  The only insecure and
not precisely known quantity is the sigma mass which influences the
phase structure~\cite{Schaefer:2009ui}. To compare with previous
works~\cite{Schaefer:2008hk} we choose $m_\sigma=600\MeV$ for the
no-sea mean-field approximation and $m_\sigma = 400\MeV$ for the
renormalized model (see also the next section).

Neglecting the Polyakov-loop potential in \Eq{eq:grandpot} and the
background gauge field $A_0$ in the covariant derivative which
corresponds to $\Phi = \bar \Phi =1$ in the quark contribution
\Eq{eq:Omegaqq} results in the corresponding three-flavor quark-meson
model. Since the Polyakov loop decouples from the QM sector in the
vacuum no parameter adjustments are necessary.

\begin{figure*}
\centering
\subfigure[$\ $No-sea approximation]{\label{sfig:pdnosea}
  \includegraphics[width=\twofigs]{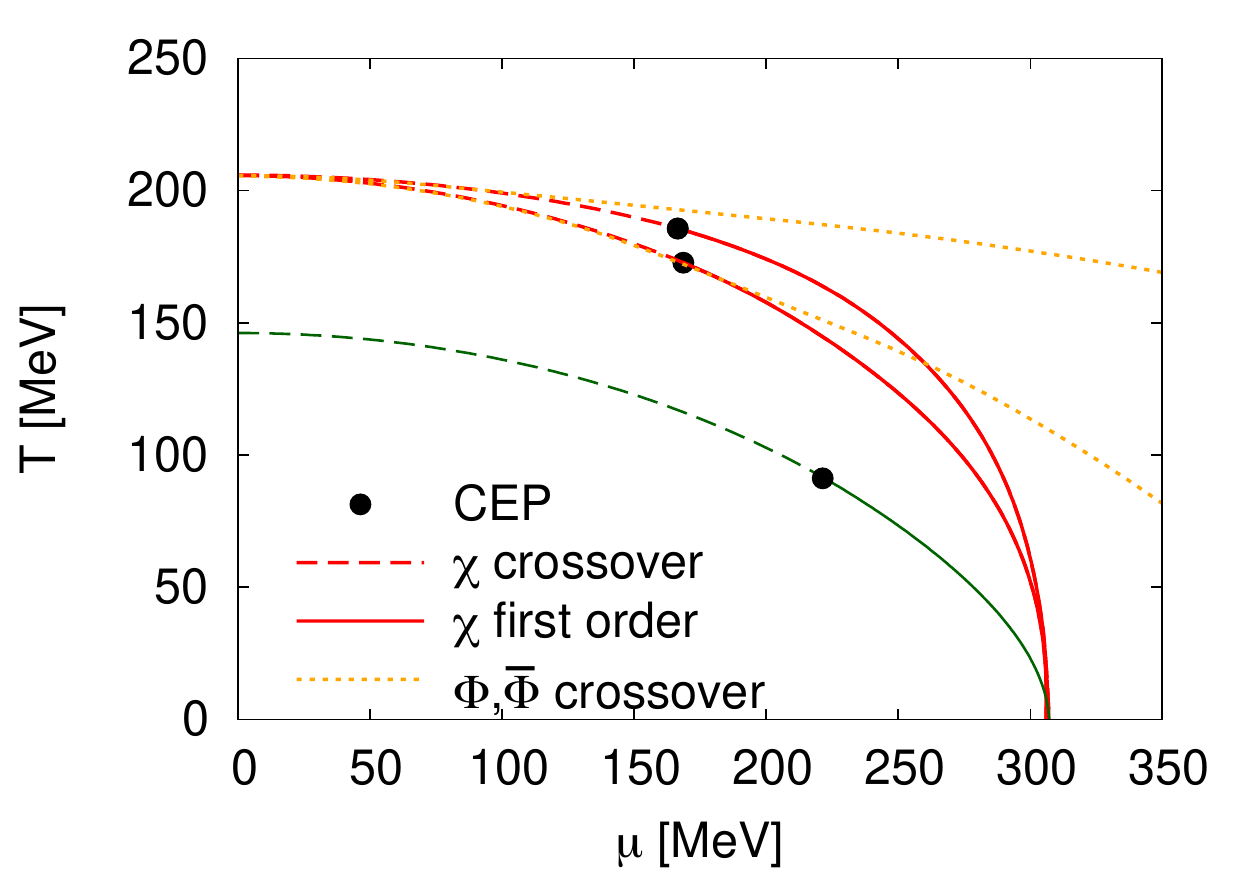}}
\subfigure[$\ $Renormalized models]{\label{sfig:pdzero}
  \includegraphics[width=\twofigs]{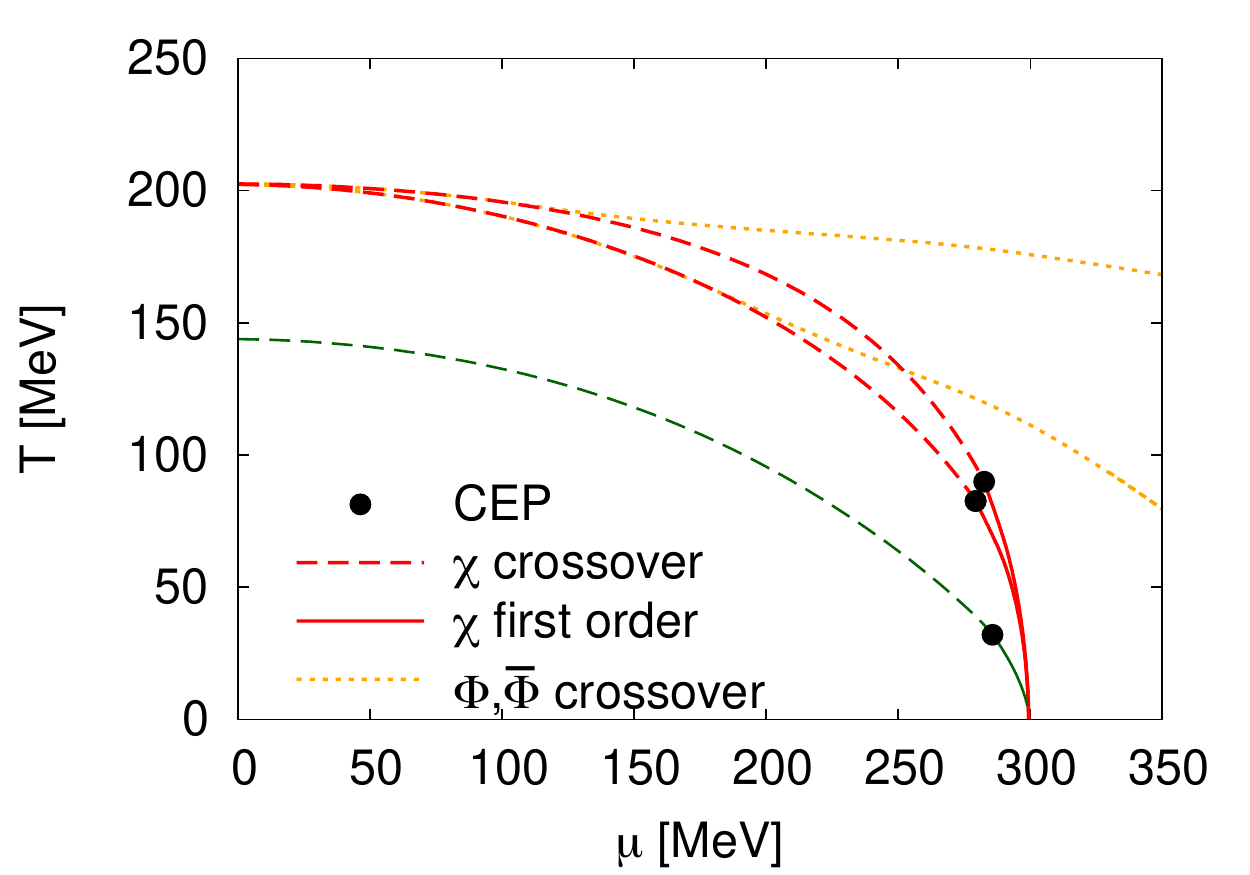}
  }
  \caption{\label{fig:pqmpd} Three-flavor chiral phase diagrams [upper
    lines, PQM models for a constant $T_0$ (top line) and for
    $T_0(\mu)$; lower lines, QM models]. (a) No-sea MFA with
    $m_\sigma=600\MeV$; (b) renormalized models with
    $m_\sigma=400\MeV$.  The peak in the temperature derivative of the
    Polyakov loops $\Phi$, $\bar \Phi$ is also shown (light dashed
    lines).}
\end{figure*}

\section{Phase structure and the critical endpoint}
\label{sec:phasestructure}

In \Fig{fig:pqmpd} the $(2+1)$-flavor chiral phase diagrams for the QM
and the PQM models with and without the matter backreaction via
$T_0(\mu)$ are presented in no-sea mean-field approximation [1(a)] and
in the renormalized models [1(b)]. The crossover in
the strange sector which takes place at higher temperatures is not
shown. The almost degenerated positions of the peaks of the
temperature derivative of both Polyakov loops $\Phi$ and $\bar \Phi$
are indicated by the light dashed lines.

In all models the vacuum term in the grand potential moves the
location of the endpoint away from the $T$ axis towards the $\mu$ axis
whereas the Polyakov loop increases the chiral transition temperature
at $\mu=0$ and hence moves the CEP back to higher temperatures.  For
example the PQM model in no-sea approximation without the matter
backreaction for a constant $T_0 = 270$ MeV yields a CEP at
$(T_c,\mu_c) = (186,167) \MeV$ which changes to $(T_c,\mu_c) =
(172,169) \MeV$ if the matter backreaction is taken into
account. Without the Polyakov loop the CEP lies at $(T_c,\mu_c) =
(92,221) \MeV$. With vacuum fluctuations the location of the CEP
changes significantly to $(T_c,\mu_c) = (90\ (83),283\ (280)) \MeV$ in
the renormalized PQM model without (with) matter backreaction. In the
renormalized QM model the CEP lies close to the $\mu$ axis at
$(T_c,\mu_c) = (32, 286) \MeV$.

The location and hence the existence of a CEP in the phase diagram is
sensitive to the sigma meson mass $m_\sigma$~\cite{Schaefer:2008hk,
  Andronic:2008gu}.  A larger $m_\sigma$ in general pushes the CEP
towards the $\mu$ axis.  Already for $m_\sigma = 600\MeV$ the CEP
disappears in the renormalized models (see also~\cite{Schaefer:2008hk,
  Chatterjee:2011jd}).  We have adapted $m_\sigma = 400$ MeV in the
renormalized models such that a common value for the chiral transition
of $T_\chi \approx 205$ MeV at $\mu=0$ is secured as in the no-sea
mean-field approximation.

In the PQM models the coincidence of the chiral and deconfinement
transitions extends to larger chemical potentials if the backreaction
to the YM sector is implemented via \Eq{eq:t0mu}.  Thus, the inclusion
of fluctuations and the matter back-coupling to the YM sector reduces
a possible chirally symmetric and confining region in the phase
diagram as found in a large-$N_c$ limit study~\cite{McLerran:2007qj}.
A similar result is obtained if one goes beyond mean field with, e.g.,
functional techniques such as the functional renormalization group \cite{Herbst:2010rf} or Dyson-Schwinger equations
\cite{Fischer:2011mz} where both transitions coincide over the whole
phase diagram.  Hence, the inclusion of the $\mu$ dependence in $T_0$
in the Polyakov-loop potential represents a further step beyond the
standard mean-field approximation. However, at vanishing temperature
all transitions collapse to the same critical $\mu$ and the
Polyakov loop does not influence the chiral sector.  Of course, in
this domain other degrees of freedom like baryons, which are usually
neglected in such (P)QM- or (P)NJL-type model studies, become more
important which we do not address here.

\begin{figure*}
  \centering
  \subfigure[$\ $ $T_0 =270\MeV$ (dashed) and with $T_0(\mu)$
  (solid)]{ \label{fig:critrregionpqmTmu}
    \includegraphics[width=\twofigs]{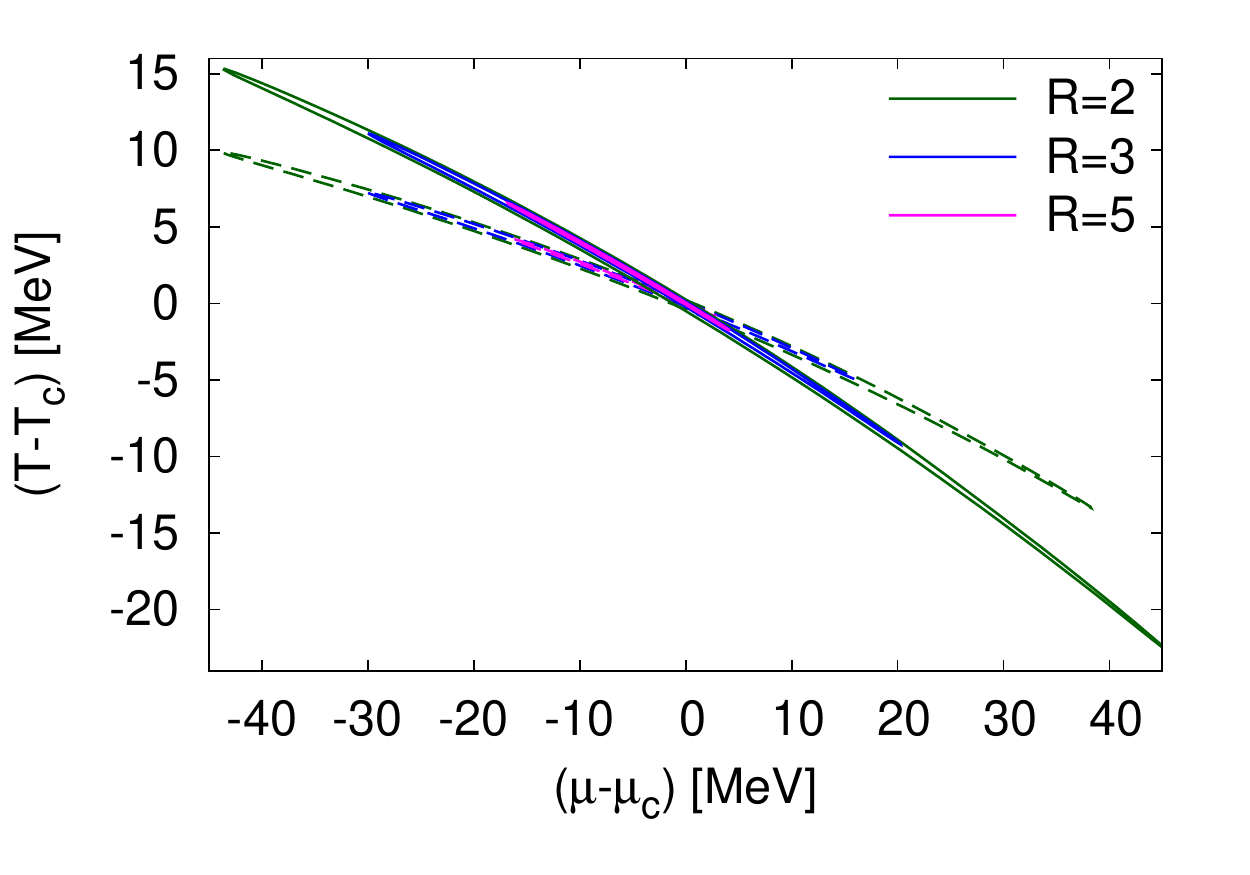}}
 \subfigure[$\ $Quark-meson model ]{ \label{fig:critrregionqm}
    \includegraphics[width=\twofigs]{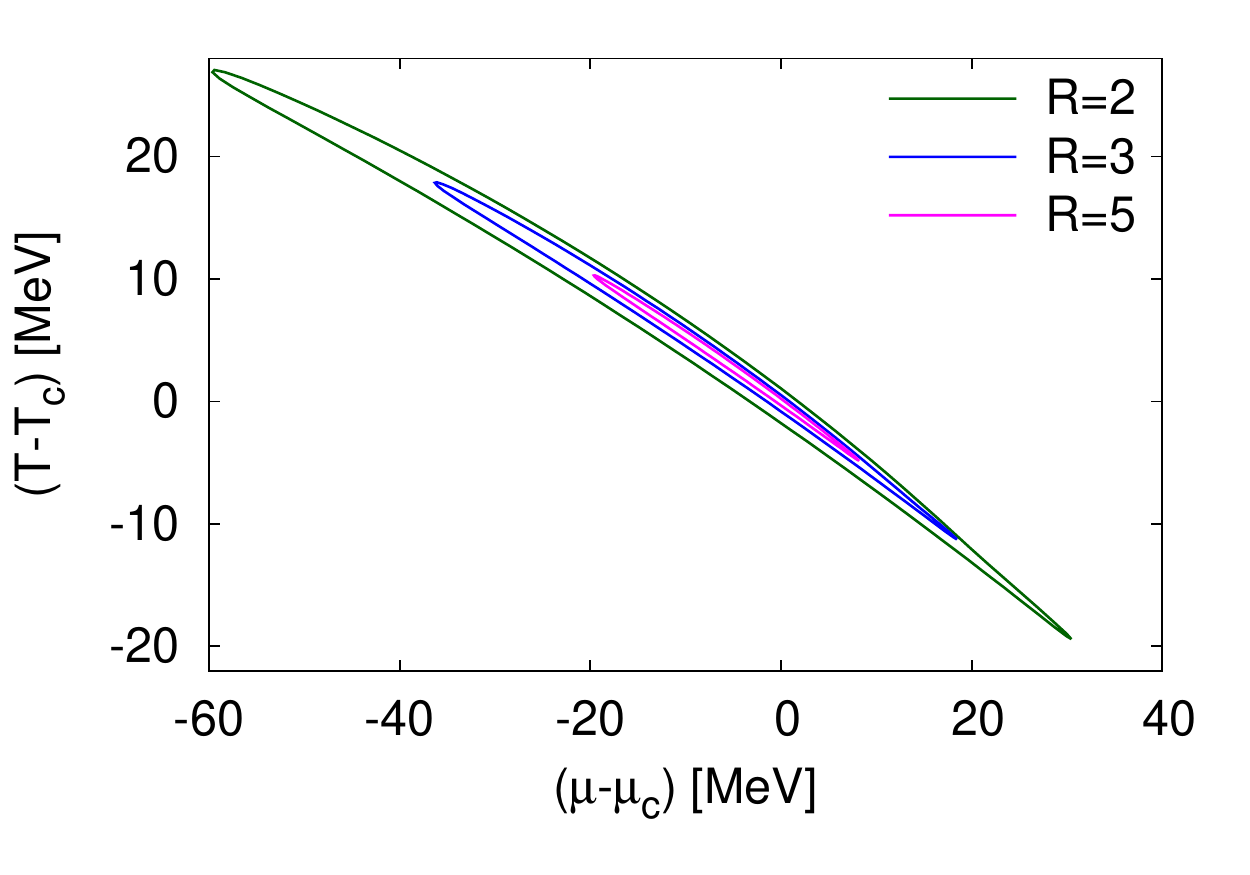}}

  \caption{\label{fig:critrregion} The critical region around the CEP
    for three different three-flavor models in no-sea MFA: PQM model
    with a logarithmic Polyakov-loop potential and constant and
    running $T_0$ (a) and the QM model (b). See text for
    details. }
\end{figure*}

\begin{figure*}
  \centering
  \subfigure[$\ $ $T_0 =270\MeV$ (dashed) and with $T_0(\mu)$
  (solid)]{\label{fig:critrregionpqmTmuVT}
    \includegraphics[width=\twofigs]{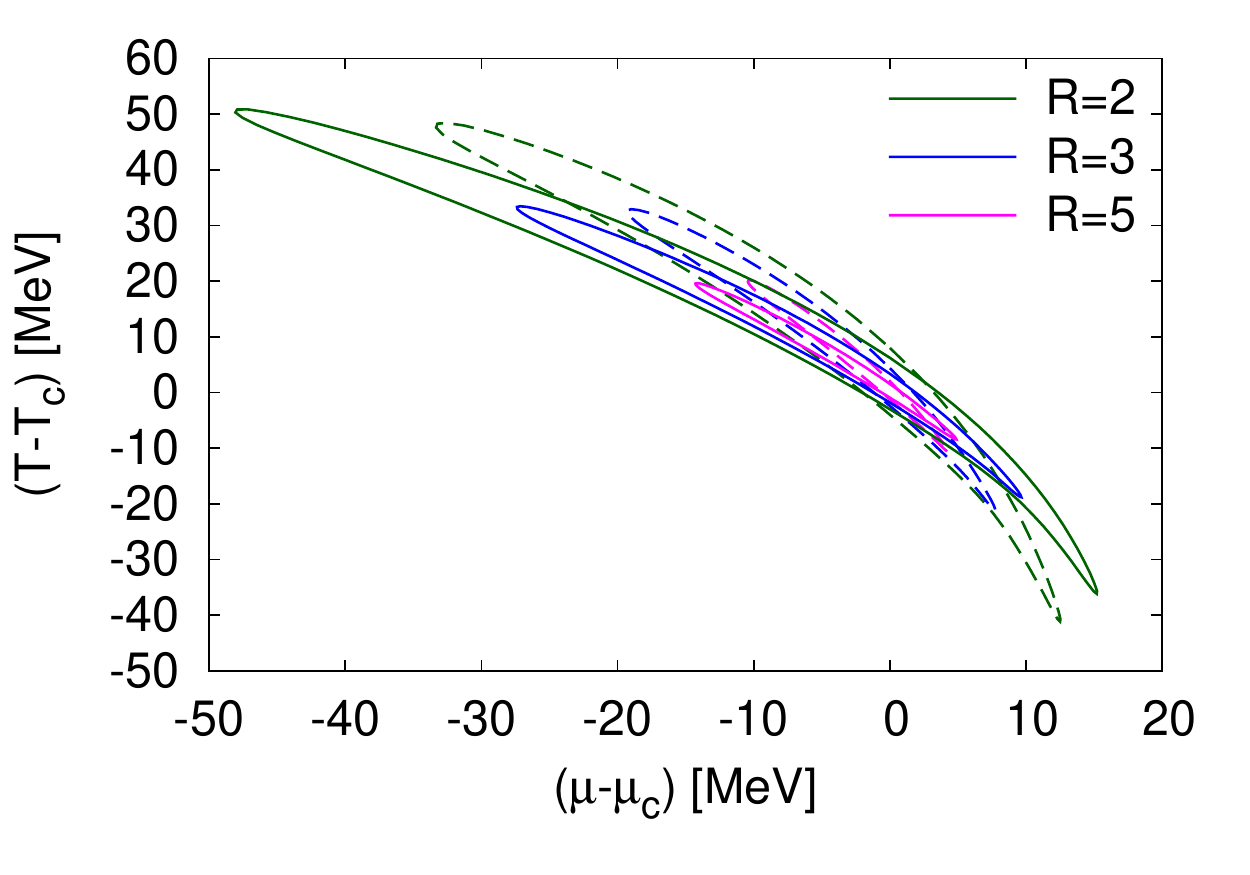}}
  \subfigure[$\ $Quark-meson model ]{\label{fig:critrregionqmVT}
    \includegraphics[width=\twofigs]{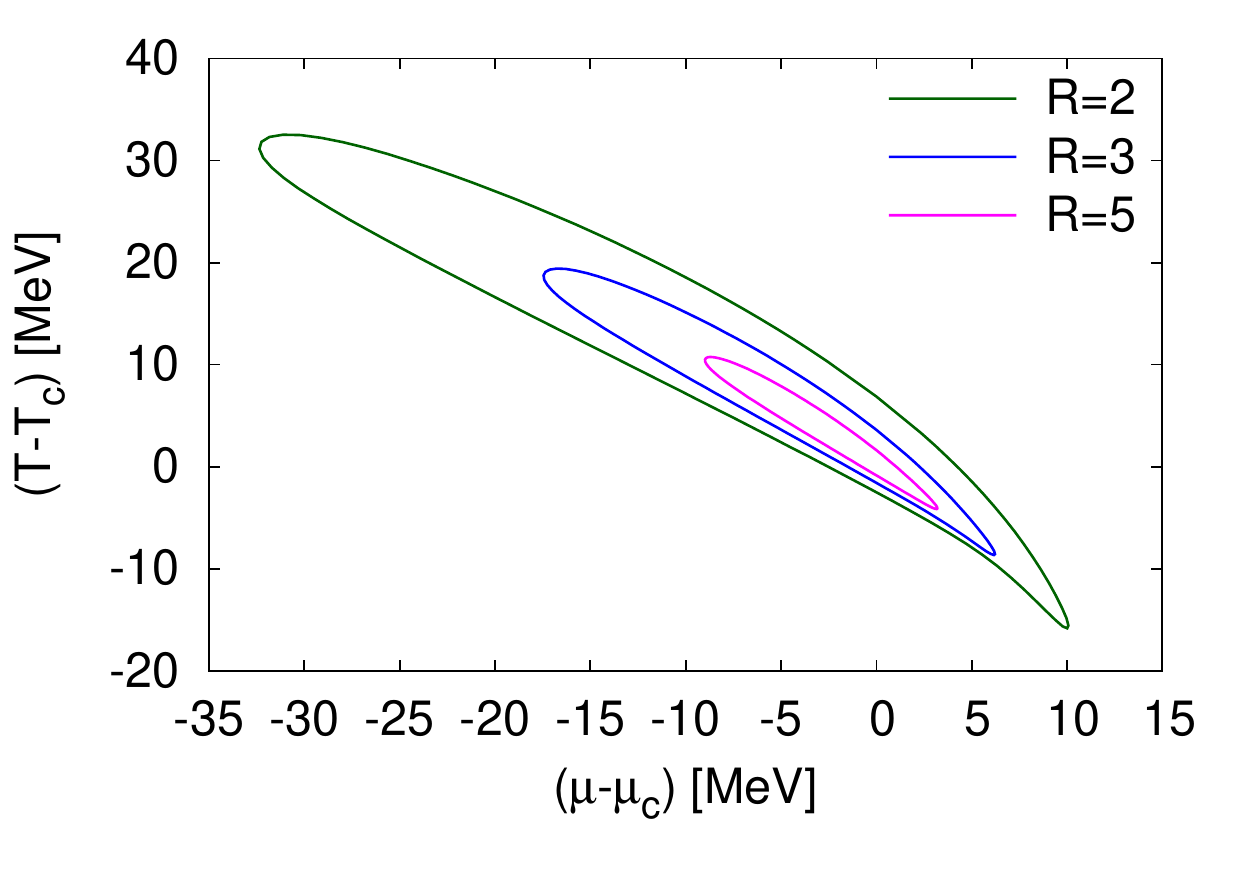}}

  \caption{\label{fig:critrregionVT}The critical region similar to
    \Fig{fig:critrregion} for the corresponding
    renormalized models.}
\end{figure*}
\subsection{Anatomy of the critical region}

The knowledge of the size of the critical region around the CEP in the
phase diagram is important for the experimental detection of the
endpoint in heavy-ion collisions. The region is obtained as a
projection of different constant ratios
\begin{equation}
  R= \frac{\chi_q}{\chi_q^{\text{free}}}
\end{equation}
of the quark-number susceptibility $\chi_q$ normalized with the free
susceptibility $\chi_q^{\text{free}}$ onto the ($T, \mu$)-plane near
the CEP. In \Fig{fig:critrregion} three ratios $R$, obtained in no-sea
MFA, are plotted as contours relative to the endpoint. Compared to the
QM model (right panel) the Polyakov loop compresses the critical
region, in particular in the $T$-direction. With the matter
back-coupling (solid lines) this effect is weaker and the region is
twice as large in the $T$-direction. In $\mu$-direction the
modification is even less pronounced.  Without the matter
back-coupling the chiral and deconfinement transition deviate already
in the vicinity of the CEP. As a consequence the quark determinant is
more suppressed near the chiral transition and hence the size of the
critical region.

The effect of the vacuum contribution in the grand potential on the
size of the critical region is shown in \Fig{fig:critrregionVT}. The
incorporated fluctuations increase the critical region in direction
perpendicular to the crossover line in the phase diagram. This effect
is more prominent in the QM model without the Polyakov loop
[3(b)]. The Polyakov loop affects the critical region less than in the
no-sea approximation. However, the region in the renormalized PQM
models [3(a)] is larger than in the renormalized QM model. This
can be understood by the observation that the Polyakov loop modifies
the quark determinant mostly at moderate chemical potentials. In the
renormalized models the CEP is located at larger chemical potentials
around $\mu\approx280-285\MeV$. Thus, the enhanced critical region
with the Polyakov loop results also from the higher critical
temperature [cf. also \Fig{sfig:pdzero}]. The matter backreaction
modifies the critical region in a similar fashion as in the no-sea
approximation.

In general, fluctuations wash out the phase transition and the chiral
crossover is much smoother. As a consequence, the size of the critical
region becomes broader in direction perpendicular to the extended
first-order transition line. But due to the shift of the endpoint away
from the $T$-axis the size of the critical region turns out to be
smaller along the phase boundary compared to the no-sea
MFA. Furthermore, for larger chemical potentials and smaller
temperatures the influence of the Polyakov loop becomes insignificant
and hence the size of the critical region becomes comparable in both
renormalized models.

All critical regions exhibit an elongation along the direction
parallel to the first-order phase transition line. The peak of the
quark-number susceptibility follows the crossover line and diverges at
the CEP.  The divergence is described by a power law within the
critical region.  The shape of the critical region is induced by the
path dependency of the corresponding critical exponents near the
singularity as already pointed out in \cite{Griffiths1970}.  One
reason for the elongation of the critical region in a direction parallel
to the first-order line is that the corresponding critical exponent is
larger \cite{Hatta:2002sj, Schaefer:2006ds}.

\begin{figure}
\centering
\includegraphics[width=0.9\linewidth]{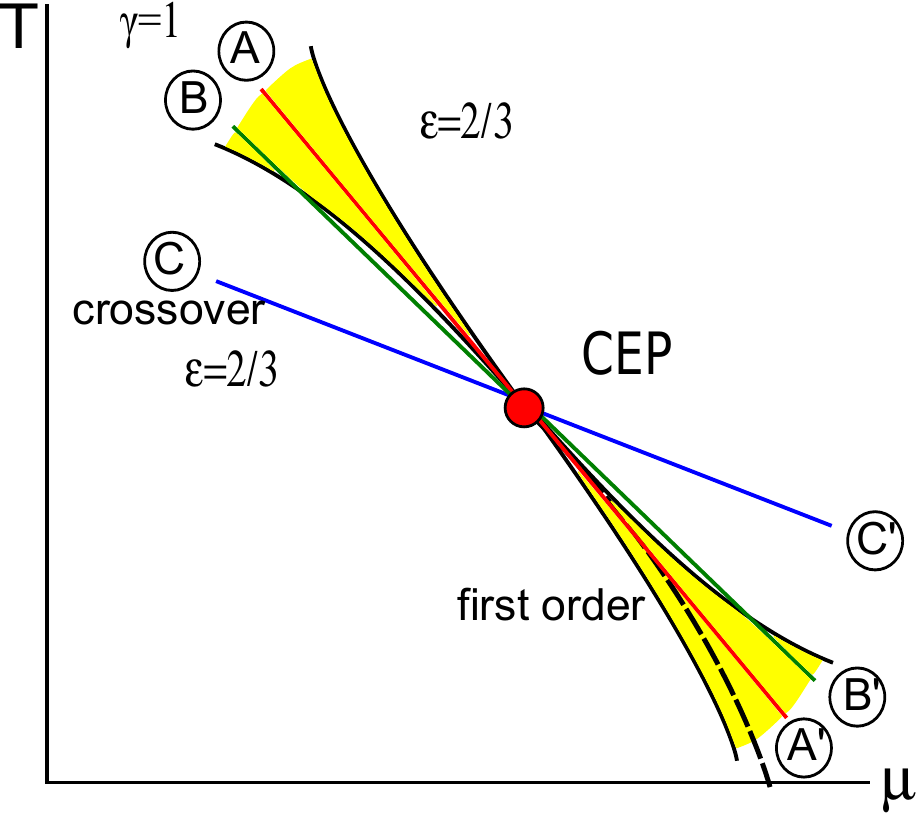}
\caption{\label{fig:scalingcrossover} Schematic plot of three
  different paths (labeled with A, B and C and with A', B' and C')
  towards the CEP in the phase diagram.  The yellow-shaded inner regions
  denote the area parallel to the (extended) first-order line (dashed)
  where the critical exponent of the quark-number susceptibility
  deviates from the one outside this area. See text for details. }
\end{figure}

\begin{figure*}
\centering
\subfigure{\label{sfig:criteps121002}\includegraphics[width=\twofigs]{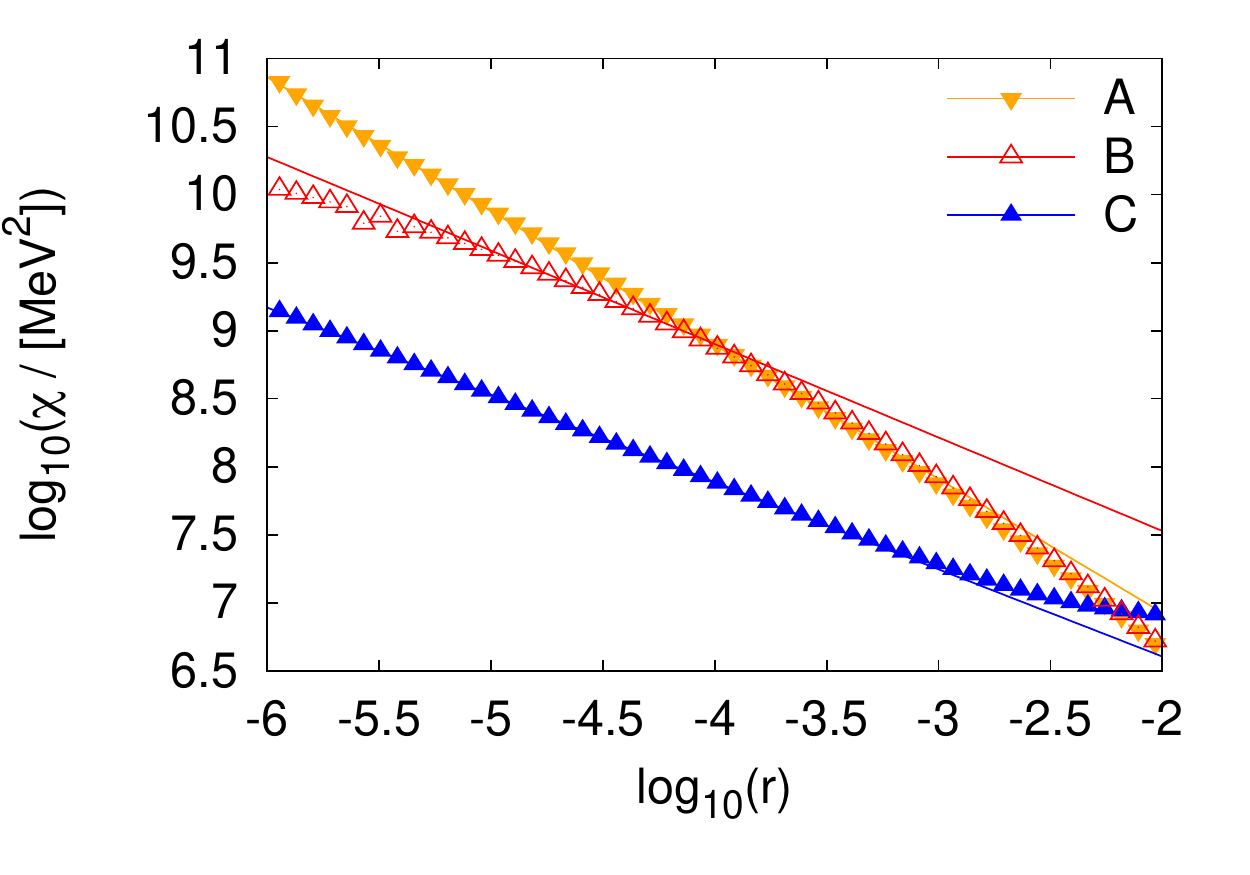}}
\subfigure{\label{sfig:criteps1002}\includegraphics[width=\twofigs]{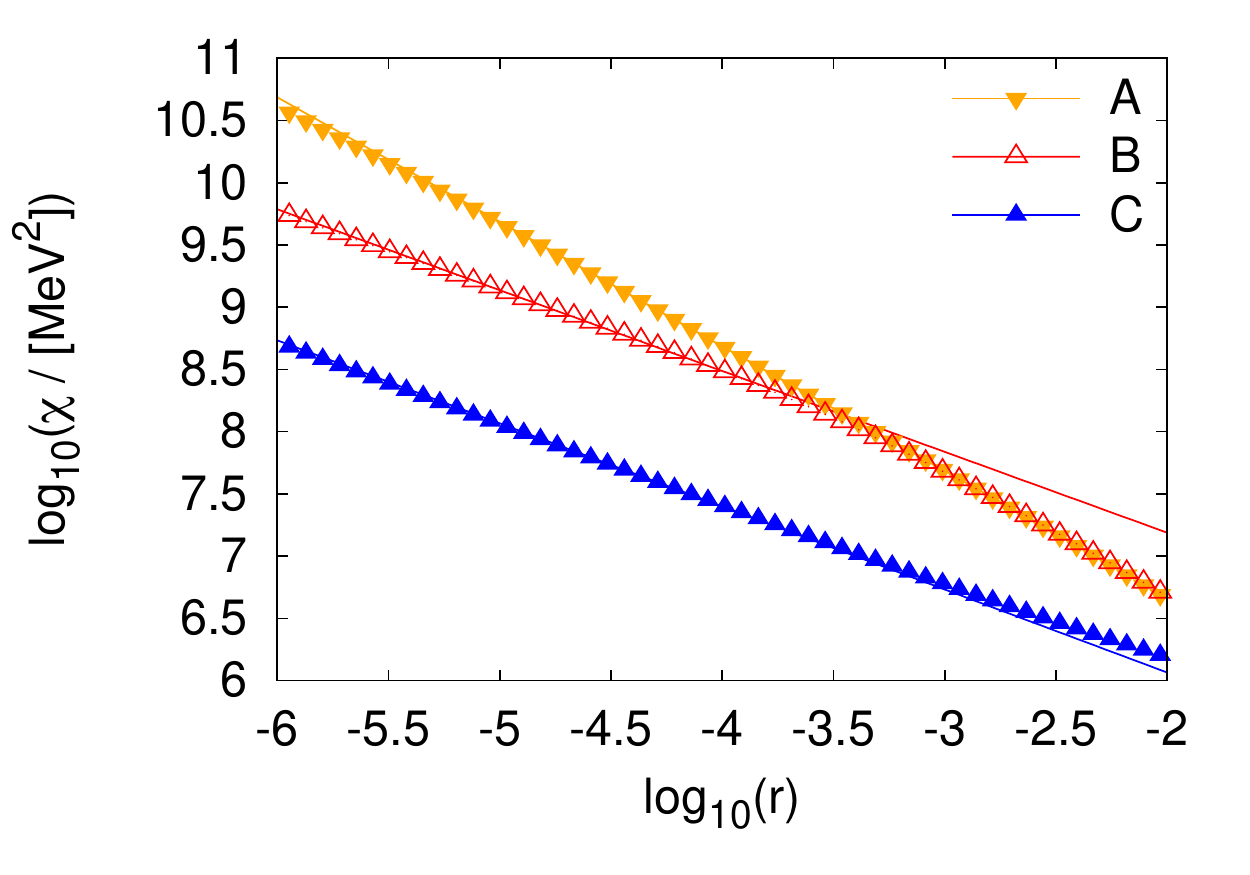}}
\caption{\label{fig:criteps}The scaling of the quark-number
  susceptibility $\chi_q$. Three different paths
  (cf.~\Fig{fig:scalingcrossover}) towards the CEP are shown for the
  PQM model (left panel) and for the QM model (right panel) in no-sea
  MFA.}
\end{figure*}

This behavior is elucidated in the following: In
\Fig{fig:scalingcrossover} a generic plot of the region near the CEP
in the phase diagram with three different paths towards the CEP,
labeled with A, B and C (A', B' and C') for $\hat t > 0 $ ($\hat t < 0
$) where $\hat{t} \equiv {(T-T_c)}/{T_c}$ is shown. Path A (A') is
chosen such that it is parallel to the (extended) first-order line.
It corresponds to the Ising model temperaturelike direction with a
vanishing "magnetic field" component in the sense of Griffiths and
Wheeler \cite{Griffiths1970}.  The remaining paths B (B')and C (C') do
have some additional "magnetic field" contributions.  For the critical
exponent $\epsilon$, which is defined by
\begin{equation}
\chi_q \propto r^{-\epsilon}
\end{equation}
with the reduced distance
\begin{equation}
r = \sqrt{\hat{t}^2 + \hat{\mu}^2}
\end{equation}
from the CEP with $\hat{\mu} = {(\mu-\mu_c)}/{\mu_c}$, two different
scaling regions can be distinguished. The explicit form and size of
these regions are not universal. They depend on the used thermodynamic
variables and the underlying equation of state; however, the mapping
is analytic \cite{Griffiths1970}.  When the $T, \mu$ variables are
used, the inner region is very small. For the outer and larger region,
perpendicular to the extended first-order line, the mean-field value
$\epsilon = 2/3$ is found.  Within a narrow area, parallel to this
first-order line, a value of $\gamma =1$ is seen which corresponds to
the Ising temperature-like exponent for vanishing external
field $\gamma = \epsilon \beta \delta = 1 > \epsilon$ (cf.~\Fig{fig:scalingcrossover}).
The critical exponent does not depend on whether one approaches the
critical point from the side with $\hat t>0$ or from $\hat t <0$.  The
scaling behavior is confirmed and demonstrated in \Fig{fig:criteps}
where the logarithm of the quark-number susceptibility for three
different paths towards the CEP (for $\hat t >0$) as a function of the
logarithm of the reduced distance $r$
is presented.

 In the figure the scaling of the susceptibility for the PQM model
 without $T_0(\mu)$ corrections (left panel) and for the QM model
 (right panel) over several orders of magnitude is shown.  The slope
 corresponds to the critical exponent. For the path B in the QM model
 one nicely recognizes the scaling crossing from the inner region with
 an exponent $\gamma = 1$ to the outer region with an exponent
 $\epsilon = 2/3$.

A similar scaling crossing in the slopes of the corresponding PQM
curves is also visible in \Fig{fig:criteps} with a small numerical
uncertainty for path B very close to the CEP.  The distance $r$ where
the crossing of the different scaling regimes in the figure takes
place depends on the chosen angle of the path relative to the CEP
coordinates. This crossing determines the boundary of the narrow
yellow-shaded regime in \Fig{fig:scalingcrossover}.
  Summarizing the findings for both models, with and without the
  Polyakov loop, the inner regime tapers with a very small angle below
  0.1 degrees.

  The criticality is also not affected by the backreaction of the
  matter fluctuations to the YM sector.  In \Tab{tab:crexponets} the
  critical exponents for paths parallel to the $\mu$ and $T$ axes
  towards the CEP with and without the $T_0(\mu)$ corrections are
  listed. We find for these paths the expected mean-field critical
  exponent $2/3$ \cite{Costa:2008gr}. For the paths parallel to the
  first-order transition line (labeled with $A$ and $A'$) an exponent
  $\gamma = 1$ is obtained. If one goes beyond the no-sea mean-field
  approximation using the renormalized potentials, the critical
  exponents remain mean field since the universality is not modified.

\begin{table}
\begin{tabular}{|c|c|c|}
\hline
 Path &  $T_0=270\MeV$ & $T_0(\mu)$\\ \hline \hline
$\hat{t} > 0$ 	&  $0.669\pm 0.002$ &$0.669 \pm 0.002$ \\
$\hat{t} < 0$ 	& $0.662 \pm 0.003$ & $0.663 \pm 0.002$\\
$\hat{\mu} > 0$		& $0.667 \pm 0.003$ &$0.669 \pm 0.002$\\
$\hat{\mu} < 0$		& $0.663 \pm 0.002$ &$0.662 \pm 0.003$\\
A' 	& $1.00 \pm 0.02$ & $1.00 \pm 0.02$ \\
A 	& $1.01 \pm 0.02$ &$1.01 \pm 0.02$\\ \hline
\end{tabular}
\caption{\label{tab:crexponets} Critical exponents of the
  quark-number susceptibility in the PQM model with a logarithmic
  Polyakov-loop potential for a constant and a running $T_0$ for different
  paths towards the CEP.
}
\end{table}

\section{Non-Gaussian Fluctuations}
\label{sec:moments}

In a realistic heavy-ion collision the detection of the critical
region, in particular of an endpoint, is hampered by a finite
correlation length as argued in the introduction.  When we approach a
critical point more and more non-Gaussian components in the
probability distribution of the corresponding order parameter are
developed.  The corresponding higher moments are much more sensitive
on a diverging correlation length and thus seem to be a more
appropriate tool to guide the experiment to locate an endpoint
\cite{Hatta:2003wn}.

Fluctuations of conserved charges are quantified by cumulants in
statistics which are directly related to generalized susceptibilities
\cite{kubo:1962, Koch:2008ia, Fu:2010ay, Cheng:2008zh}. They can be
defined as derivatives of the logarithm of the partition function with
respect to the appropriate chemical potentials. For three different
chemical potentials we have accordingly
\begin{equation}\label{eq:defchi}
\chi_{n_i,n_j,n_k} \equiv \frac{ 1}{VT} \frac{ \partial^{n_i}}{\partial (\mu_{i}/T)^{n_i}}
\frac{ \partial^{n_j}}{\partial (\mu_{j}/T)^{n_j}}
\frac{ \partial^{n_k}}{\partial (\mu_{k}/T)^{n_k}} \log Z \ ,
\end{equation}
where $n_i,\ldots$ denotes the order of the derivatives and the
indices $(i,j,k) = (u,d,s)$ the quark flavor.

The generalized susceptibilities evaluated at vanishing $\mu_f$ are
the Taylor expansion coefficients of the pressure series in powers of
$\mu_f/T$. The coefficients and the convergence properties of the
Taylor series were recently analyzed in a similar model
study~\cite{Karsch:2010ck, Karsch:2010hm, *Karsch:2011yq}.  Moreover,
higher-order susceptibilities are also known up to eighth order from
QCD lattice simulations \cite{Gavai:2004sd, *Gavai:2008zr,
  Allton2005gk}.

For the corresponding $n$th order moment of the quark-number
fluctuations for one uniform quark chemical potential, we write
$\chi_n$.  The ratio of different order susceptibilities is volume
independent and can thus be directly linked to moment products of
various baryon number distributions. Examples are the net-proton
kurtosis $\kappa$ and the variance $\sigma$ which can be combined to
$\kappa \sigma^2 = \chi_4/(9\chi_2)$ and the net-proton skewness $S$
which can be expressed as $S\sigma = \chi_3/(3\chi_2)$. The factors
arise because $\chi_n$ are the quark-number susceptibilities. These
moments establish a deeper connection between theoretical predictions
and experimental measurements: only in the vicinity of the CEP might the
products $\kappa \sigma^2$ and $S\sigma$ show large deviations
from its Poisson value. In the crossover regime at smaller chemical
potentials, they are close to 1. Presently, these quantities are
the subject of ongoing discussions \cite{Asakawa:2009aj,
  *Redlich:2010gx, Gavai:2010zn,
  Mukherjee:2011td}.  Unfortunately, the existing statistics on the
experimental side are not yet sufficient to draw any reliable
conclusions which might point to the existence of a critical point in
the phase diagram \cite{Kumar:2011de}.

In the following we denote the $n$th to $m$th order moment ratio as
\begin{equation}
  R_{n,m}\equiv \chi_n (T,\mu)/\chi_m (T,\mu)\ ,
\end{equation}
such that, e.g., $\kappa \sigma/S = R_{4,3}/3$. The direct model
evaluation of $R_{n,m}$ becomes quickly cumbersome and tedious already
for lower orders because the order parameters depend implicitly on the
chemical potential.

The evaluation of the ratios can be automated by means of a novel
numerical derivative technique, based on the algorithmic
differentiation (AD) idea.  With this technique derivatives of
arbitrary order at machine precision can be obtained. Compared to
other differentiation techniques such as the standard divided
differentiation method or symbolic differentiation, the AD is
faster and produces truncation-error-free derivatives of a given
function.  Details and a comprehensive introduction to the AD
technique is given in \cite{GrWa08}.  Physics applications of this
technique can be found at various places: The AD generalization to
implicitly dependent functions which is also needed in this work was
done in \cite{Wagner:2009pm} and in \cite{Schaefer:2009st} the Taylor
coefficients to very high orders for a three-flavor model have been
calculated for the first time at vanishing chemical potential. These
findings confirm impressively the power and usability of this
technique. In this work we extend this method to nonvanishing
chemical potentials.

\begin{figure}
\centering
 \includegraphics[width=\onefig]{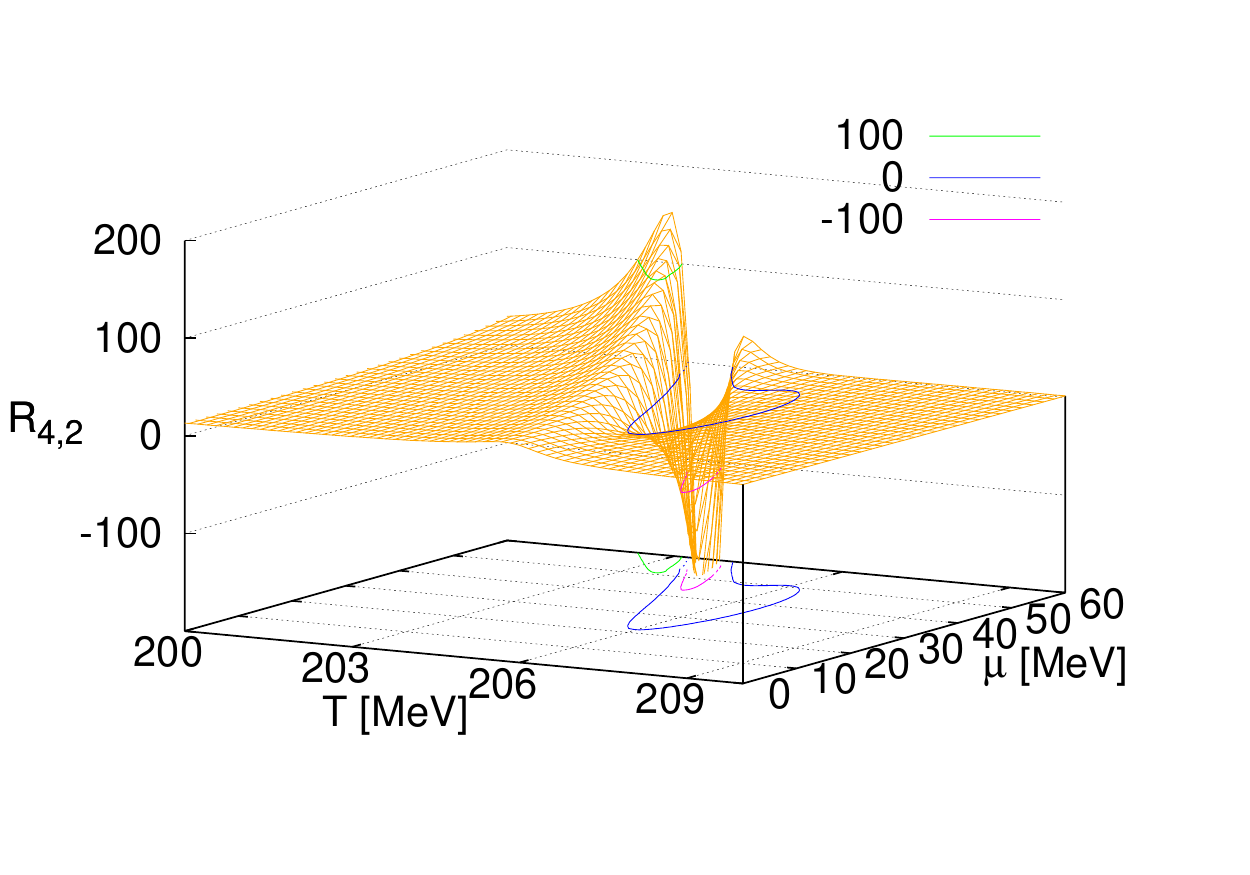}
 \caption{\label{fig:r423d}The ratio $R_{4,2}$ in the
   PQM model with a constant $T_0$.  }
\end{figure}

\begin{figure*}
  \centering \subfigure[\label{fig:negpqmnosea}$\ $Polyakov-quark-meson model with running
  $T_0(\mu)$]{
    \includegraphics[width=\twofigs]{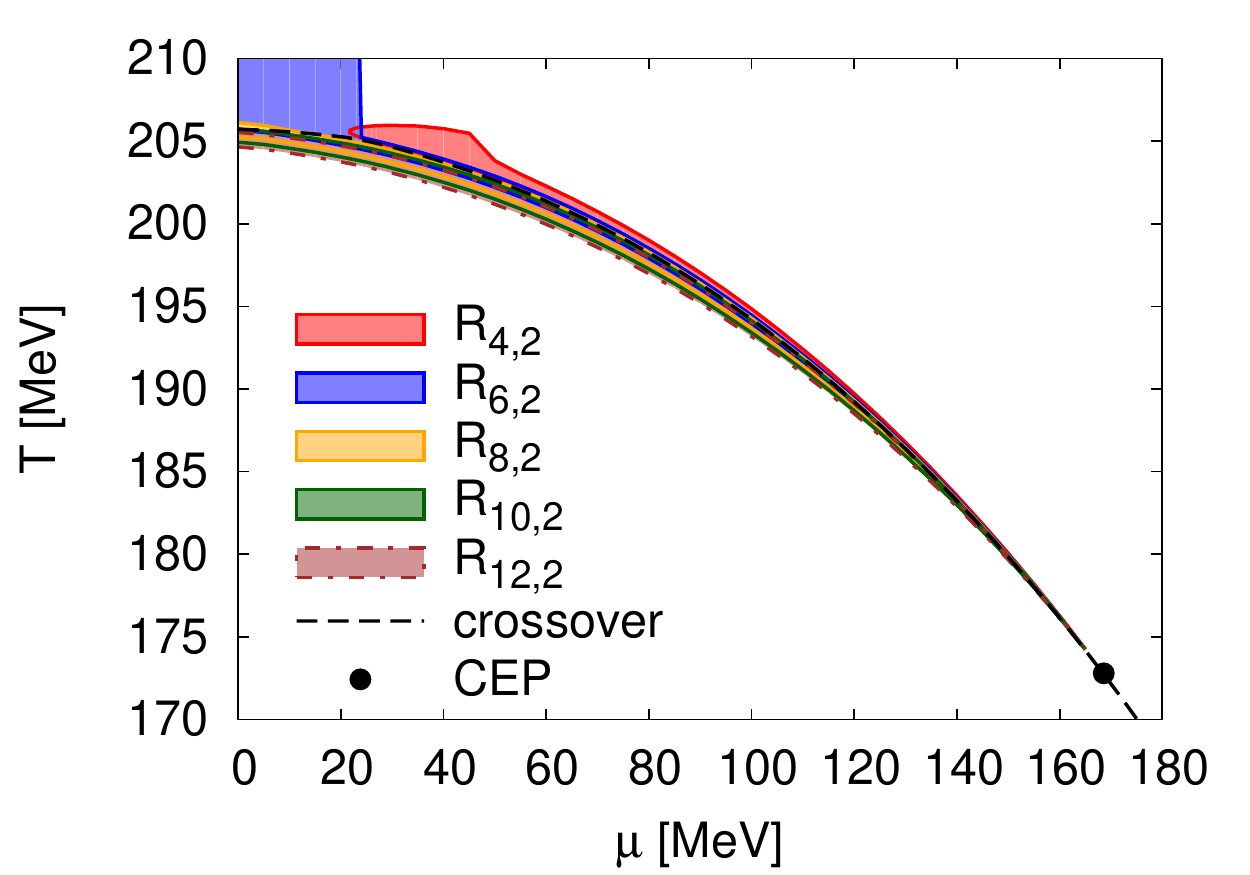}
  } \subfigure[$\ $Quark-meson model]{
    \includegraphics[width=\twofigs]{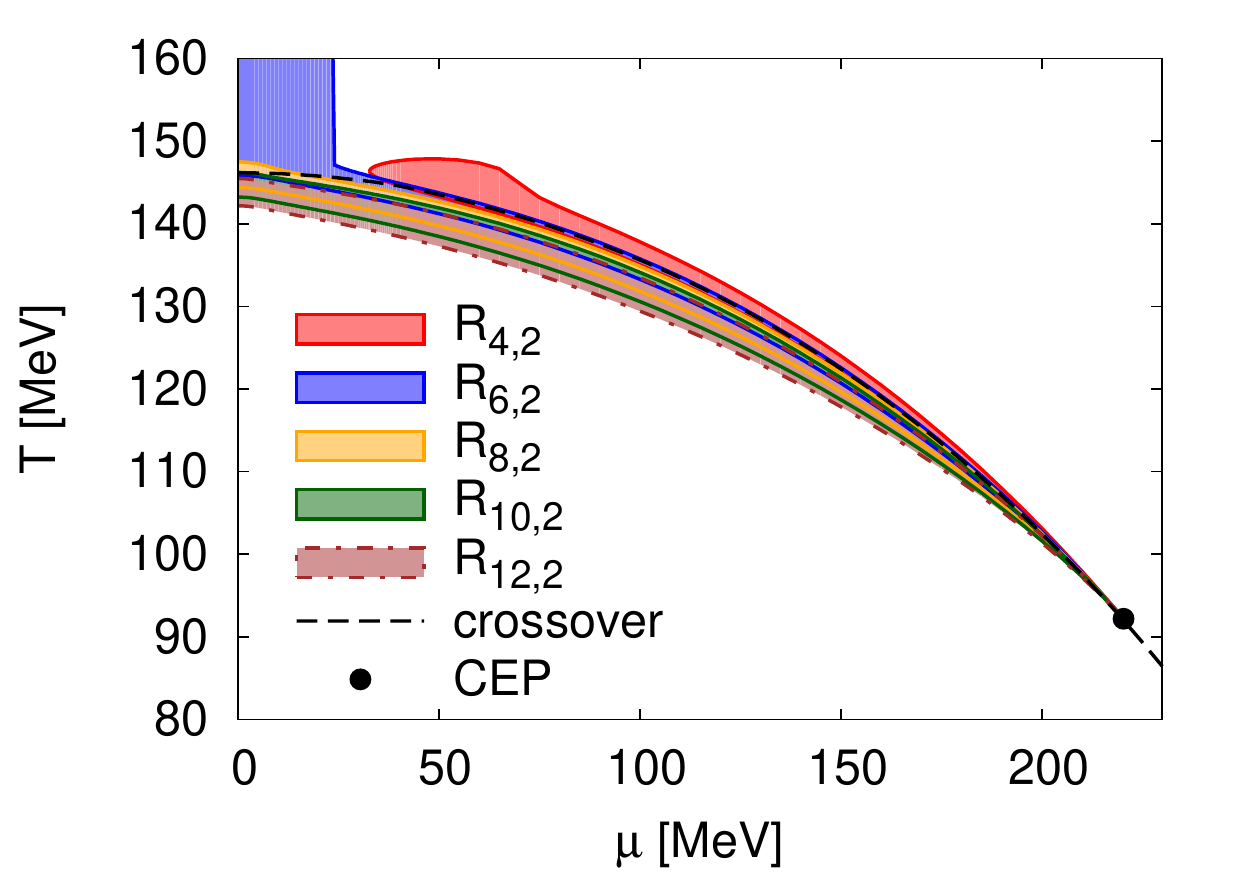}
  }
  \caption{\label{fig:extremapd} Negative regions of the ratios
    $R_{n,2}$ along the crossover line in the phase diagram. }
\end{figure*}

\begin{figure*}
  \centering \subfigure[$\ $Polyakov-quark-meson model
  with running $T_0(\mu)$]{
    \includegraphics[width=\twofigs]{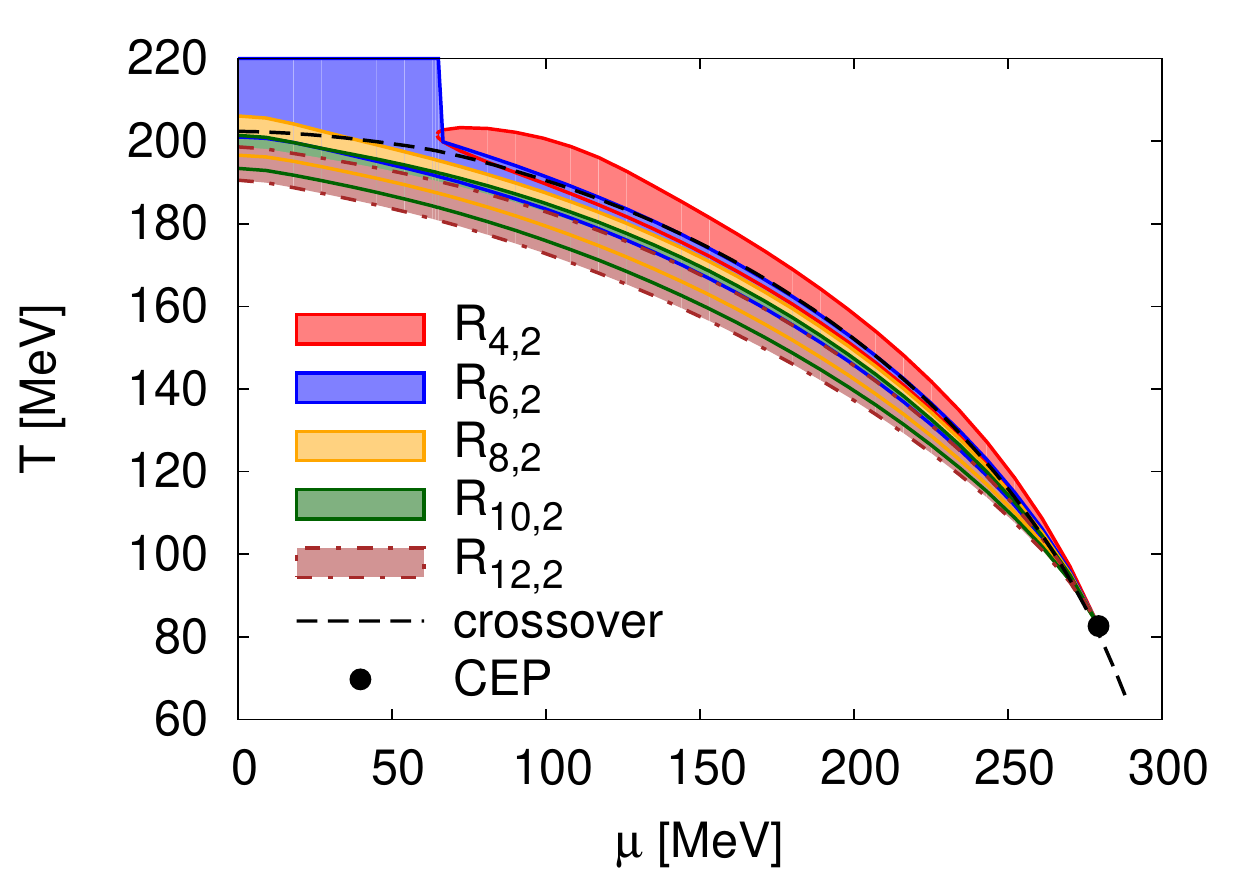}
  } \subfigure[$\ $Quark-meson model]{
    \includegraphics[width=\twofigs]{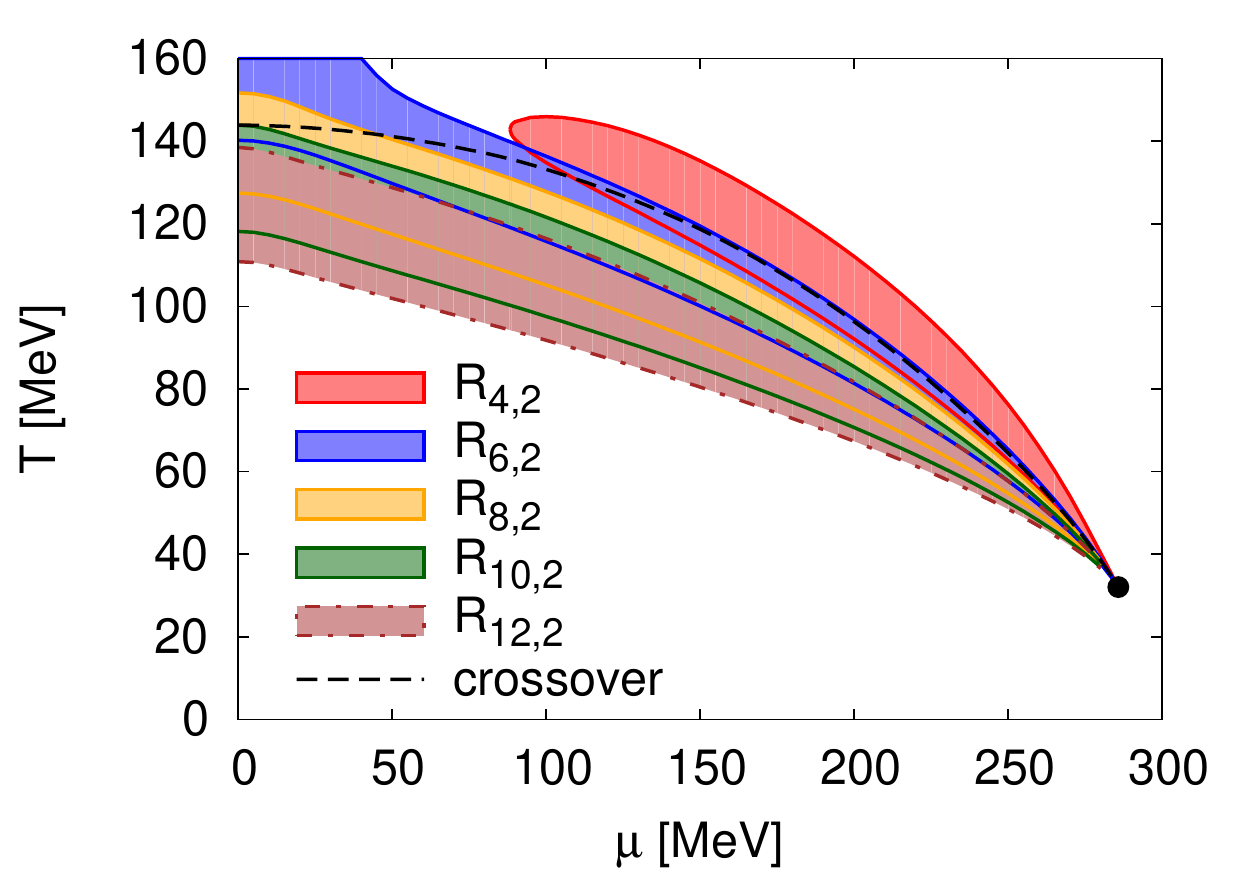}
  }
  \caption{\label{fig:extremapdVT} Negative regions similar to
    \Fig{fig:extremapd} for the corresponding renormalized models.}
\end{figure*}

We begin with a discussion of the ratios $R_{n,2}$ with even
$n\geq4$. Since the moment $\chi_2$, i.e., the quark-number
susceptibility, is always positive and continuous in the chiral
crossover regime it cannot modify the sign structure of $R_{n,2}$ near
the phase transition.  Thus, the ratios $R_{n,2}$ are a suitable
quantity to inspect the change of sign of all higher-order moments
$\chi_n$ in the phase diagram.

The ratio $R_{4,2}$ measures basically the quark content of particles
carrying baryon number.  For vanishing quark chemical potential this
ratio (if quark-number susceptibilities are used) tends to nine for
low temperature and for high temperature to $6/\pi^2$ which follows
immediately from the Stefan-Boltzmann limit.  In a hadron resonance
gas (HRG) model the ratio is temperature independent and all moments
stay positive since the HRG has no phase transition with
singularities. Hence, higher-order moments can differ significantly
from a HRG calculation along the chemical freeze-out line even if the
lower-order moments agree with HRG results and the thermodynamics and
the particle yields at low temperature are well described by the HRG
model \cite{Gavai:2010zn, Friman:2011pf, Andronic:2008gu}.  Any
deviation from the HRG model result might be an indicator for a real
critical phenomenon (in equilibrium) and thus serves as a theoretical
baseline for the analysis of heavy-ion collisions \cite{Gavai:2010zn,
  Karsch:2010ck}.

\begin{figure*}
  \centering
  \subfigure[$\ $Polyakov-quark-meson model with running
  $T_0(\mu)$]{\includegraphics[width=\twofigs]{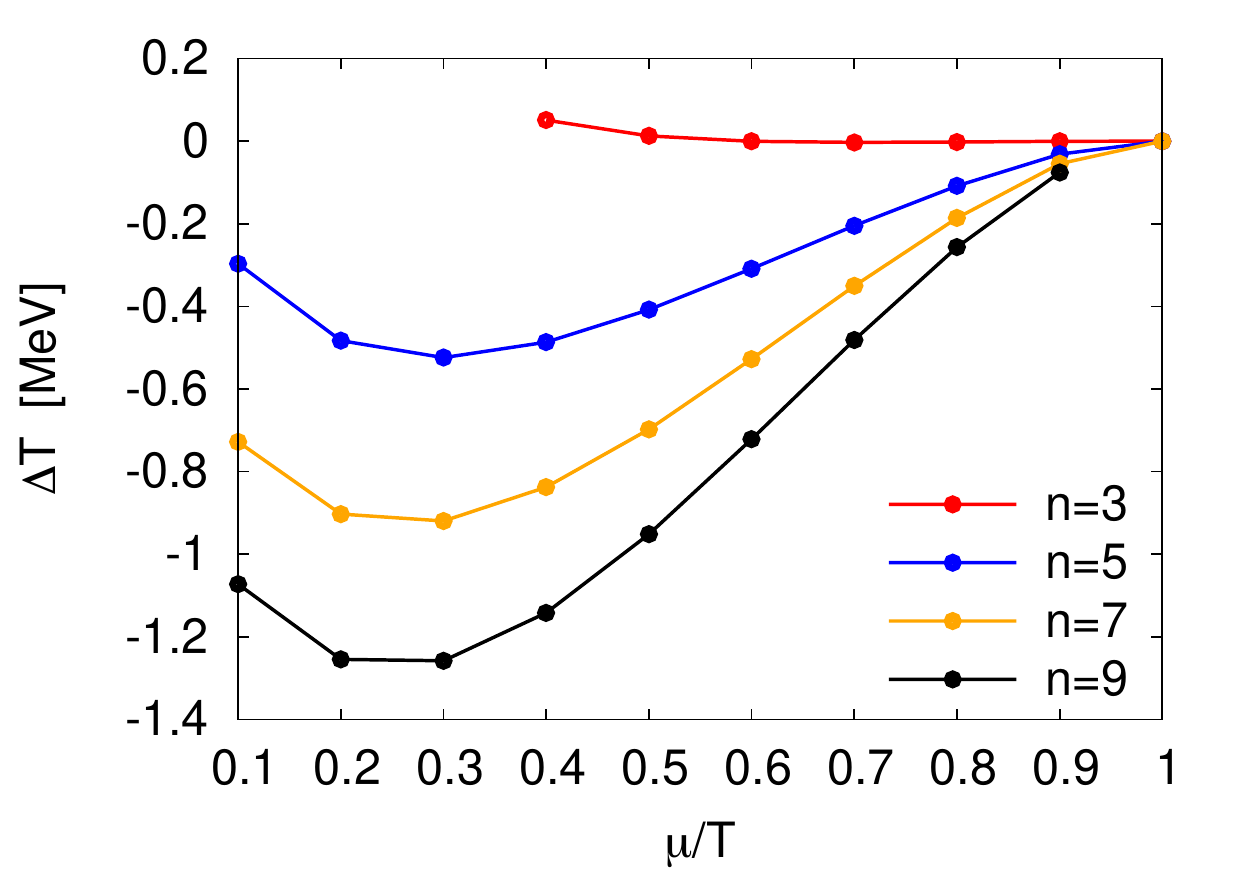}}
  \subfigure[$\ $Quark-meson model]{\includegraphics[width=\twofigs]{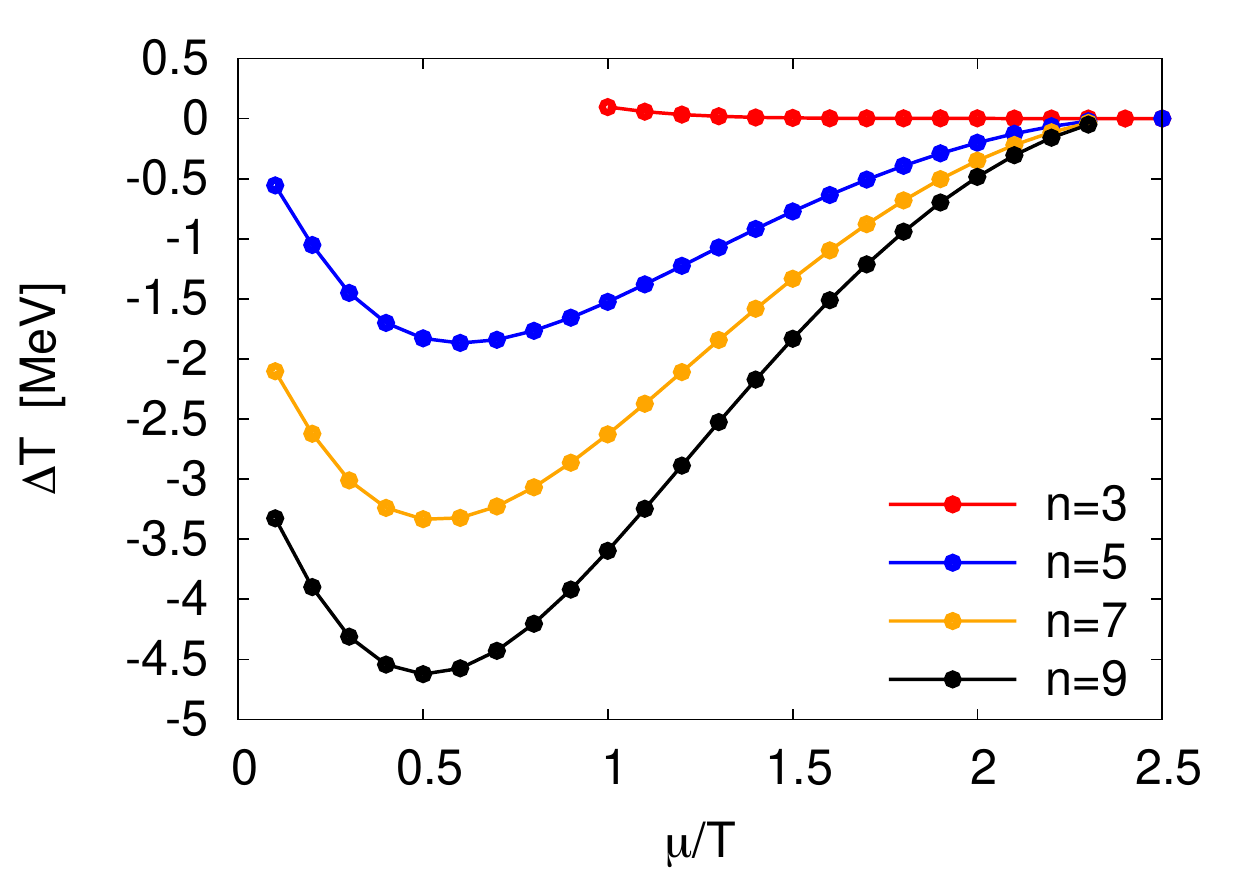}}
  \caption{\label{fig:oddroots_pb} Distance $\Delta T \equiv T_n - T_\chi$
    of the first zero for various ratios $R_{n,2}$ to the chiral
    temperature $T_\chi$ as a function of $\mu/T$ in no-sea MFA. }
\end{figure*}

\begin{figure*}
  \centering
  \subfigure[$\ $Polyakov-quark-meson model with running $T_0(\mu)$]{ \includegraphics[width=\twofigs]
   {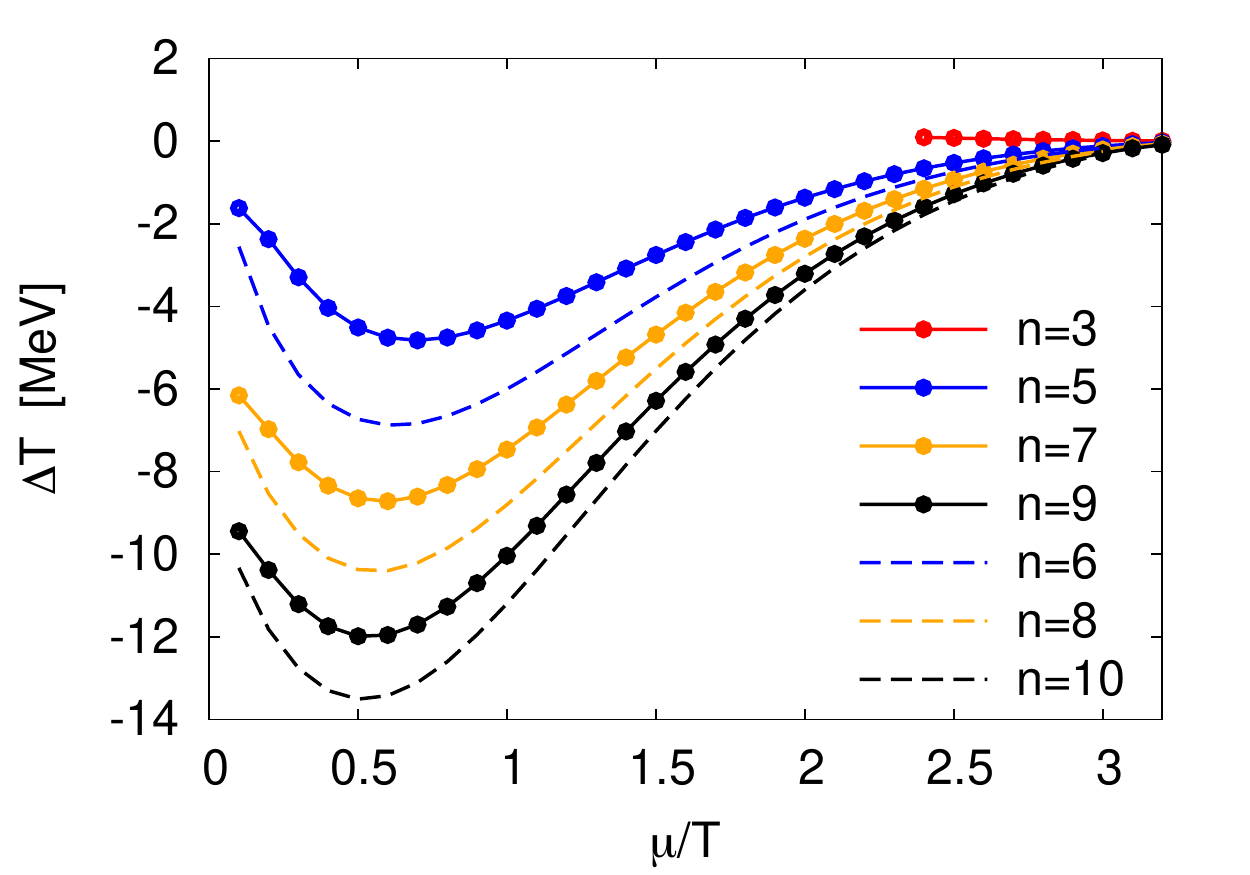}}
     \subfigure[$\ $Quark-meson  model]{ \includegraphics[width=\twofigs]{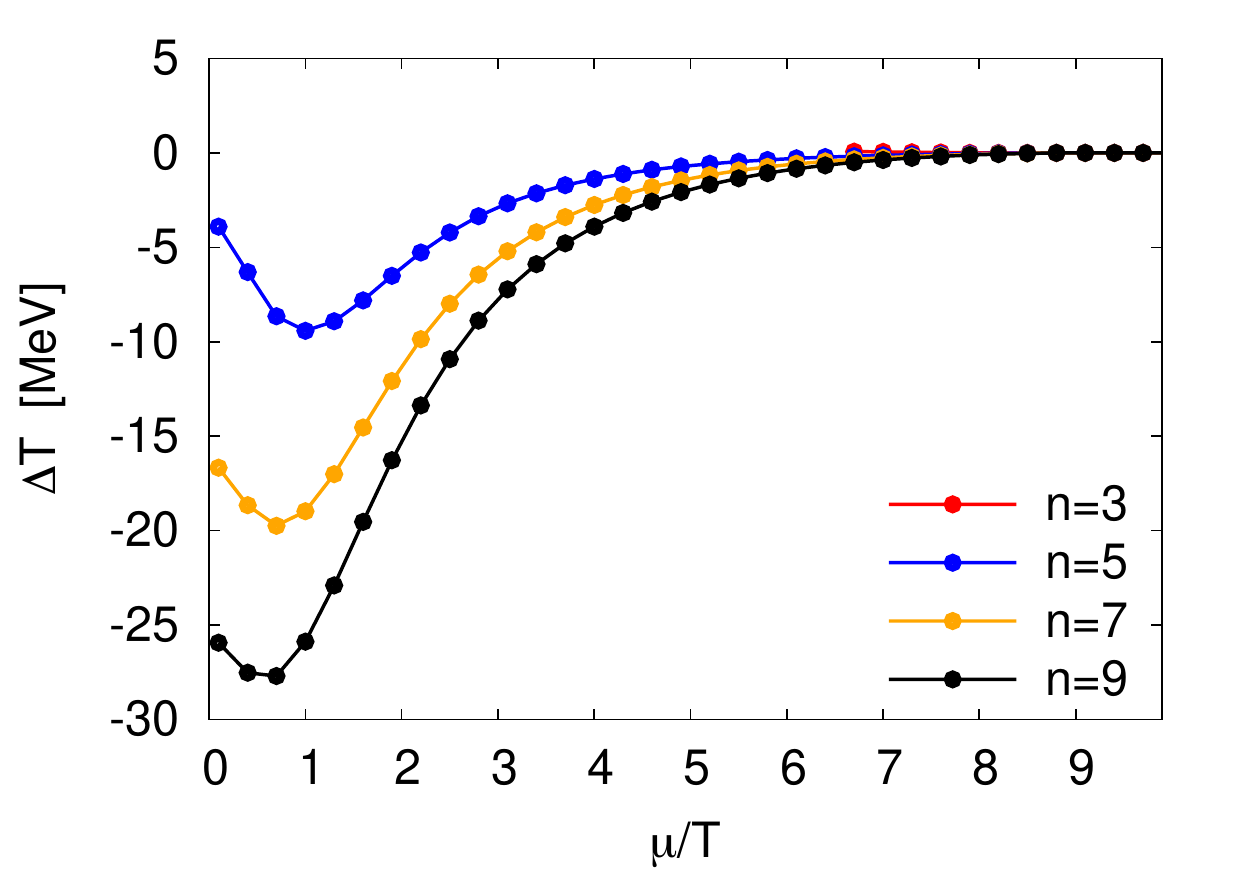}}
  \caption{\label{fig:oddroots_pbVT} Distance $\Delta T$ similar to
    \Fig{fig:oddroots_pb} for the corresponding renormalized models.}
\end{figure*}

In \Fig{fig:r423d} the ratio $R_{4,2}$ calculated in the PQM model in
no-sea MFA with a constant $T_0$ around the crossover region is
shown. The contour lines projected onto the temperature and chemical
potential plane are close to the chiral crossover line in the phase
diagram.  One clearly recognizes the development of a negative region
for nonvanishing chemical potentials.  These three-flavor results
differ from a two-flavor mean-field calculation with and without the
vacuum term \cite{Skokov:2010sf}. Without the vacuum term the
two-flavor kurtosis develops a peak at $\mu=0$ near the crossover
which is a remnant of the first-order transition in the chiral limit
\cite{Nakano:2009ps}.  The fermion vacuum term changes the
transition to second-order in the chiral limit consistent with
universality arguments \cite{Schaefer:2004en, Pisarski:1983ms}. For
three massless flavor a first-order transition always emerges in the
chiral limit with or without the vacuum term. The peak of the kurtosis for physical
pion masses is already less pronounced in the no-sea mean-field
approximation. This suppression at $\mu=0$ is caused by the strange
quarks and not by fluctuations (cf.~Fig.~7 in \cite{Schaefer:2009ui}).
Including the vacuum term the peak in the kurtosis finally vanishes
also for the three-flavor case.

The higher moments start to oscillate within a narrow temperature
interval for temperatures near the chiral transition
\cite{Schaefer:2009st}.  The structures of all moments at $\mu=0$ are
related to each other and the behavior including the amplitudes of
$\chi_n$ can be estimated by the temperature derivative of
$\chi_{n-2}$ \cite{Karsch:2010hm}. In contrast to the HRG model the
ratios in our calculation become negative in the vicinity of the
crossover for nonvanishing chemical potential as shown in
\Fig{fig:r423d}. Since the fourth-order moment $\chi_4$ is unaffected
by chiral critical phenomena at vanishing chemical potential it
remains positive for $\mu <20$ MeV. This changes for larger chemical
potentials since then nontrivial contributions from higher-order
moments evaluated at $\mu=0$ emerge and induce the change of sign in
the ratios \cite{Schmidt:2010xm, Cheng:2008zh}. This can be seen
explicitly by a Taylor expansion of the corresponding moment around
$\mu=0$.

Higher-order ratios behave in a similar way \cite{Cheng:2008zh,
  Schaefer:2009st}. In \Fig{fig:extremapd} the negative regions of
several ratios $R_{n,2}$ along the chiral crossover line are compared
to each other for two models in no-sea MFA. All negative regions are
closely correlated with the crossover curve. For ratios with $n>4$ the
negative regions are shifted slightly away from the crossover curve in
the hadronic phase and all regions converge exactly at the CEP. The
Polyakov loop sharpens the size of the negative region in the phase
diagram.

The effect of the vacuum fluctuations on the negative regions is
demonstrated in \Fig{fig:extremapdVT} where similar to the previous
\Fig{fig:extremapd} the results of various even $R_{n,2}$ ratios with
the corresponding renormalized models are shown. The CEP is pushed
towards higher chemical potentials and the crossover is washed out by
fluctuations. The curvature of the crossover line seems not to be
changed. The effect of the fluctuations is drastic in the renormalized
quark-meson model [8(b)] but not so large when the
Polyakov loop is considered [8(a)].

We conclude that the negative fluctuations can surely be attributed to
critical dynamics. Furthermore, these findings underline once more the
importance of fluctuations. Generically, one observes that all regions
calculated in the renormalized models are shifted more in the hadronic
phase.

The appearance of negative values in $\chi_n$ at $\mu=0$ has already
been suggested as a criterion to determine the chiral critical
temperature with Taylor expansion methods in lattice simulations. For
example, the $\mu=0$ values of \Fig{fig:negpqmnosea} correspond to
the ones in Fig.~5 of Ref.~\cite{Karsch:2010hm}. However, the critical
temperature estimated with this criterion deviates strongly from the
chiral critical one.  The knowledge how the negative regions evolve
towards the endpoint might be used to construct new criteria to
improve the $T_c$ estimate from a Taylor expansion around $\mu=0$.

\begin{figure*}
  \centering \subfigure[$\ $Without vacuum
  contribution]{\includegraphics[width=\twofigs]{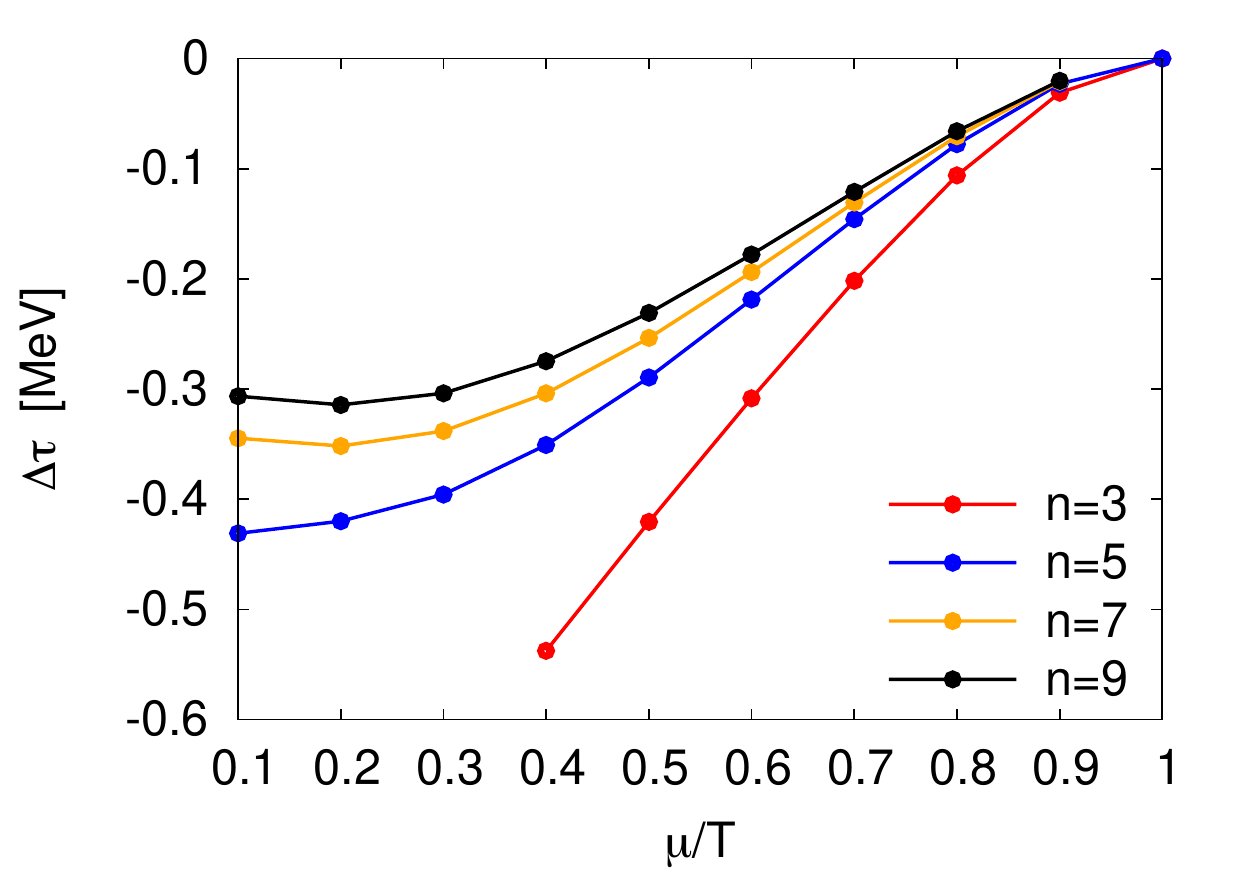}}
 \subfigure[$\ $With vacuum contribution]{\includegraphics[width=\twofigs]{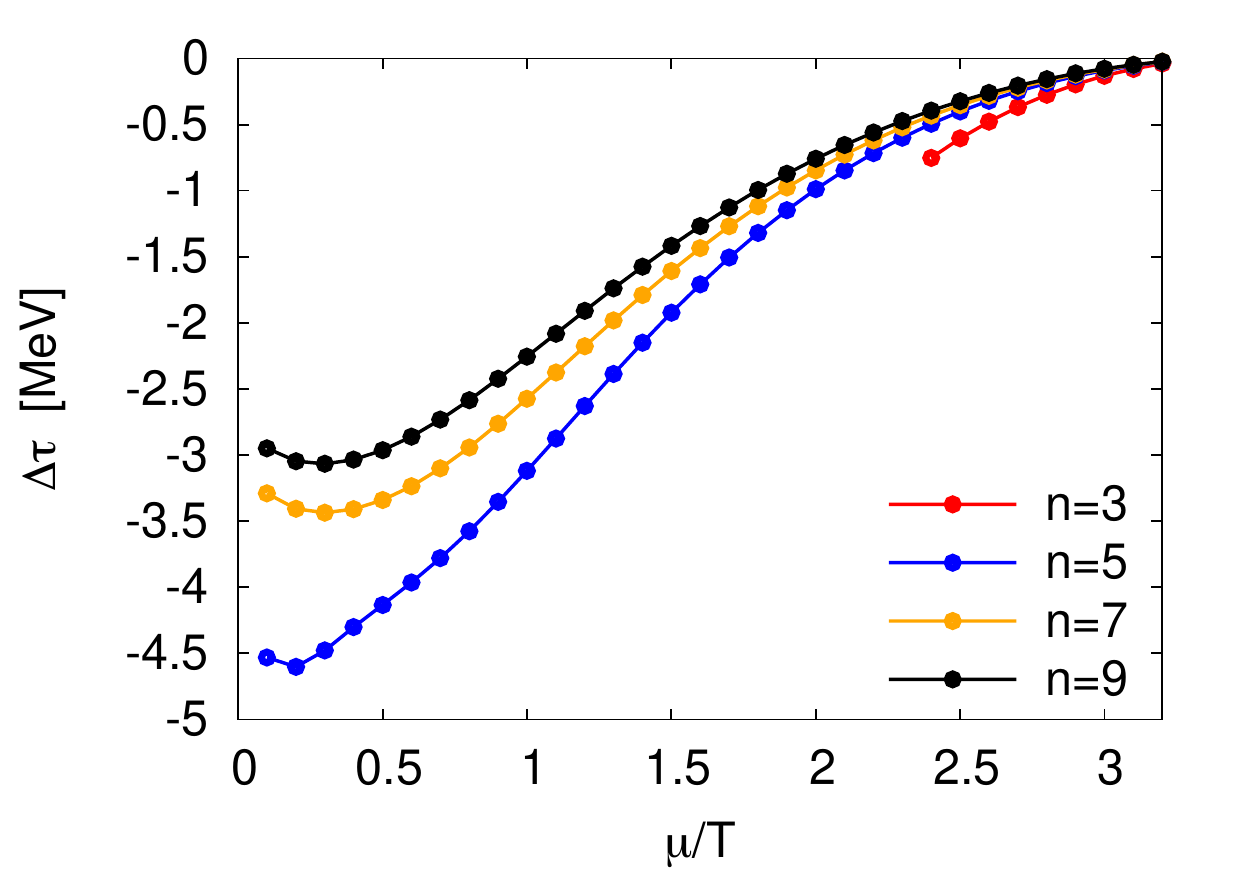}}
  \caption{\label{fig:oddrootsroots} Relative temperature distance
    $\Delta \tau \equiv T_{n+2}- T_n$ of the first zero in various
    subsequent odd ratios $R_{n,2}$ for the PQM model with running $T_0(\mu)$.}
\end{figure*}

We close this section with a discussion of $R_{n,2}$ for odd integers
$n$ which have not been addressed in the literature so far.  They
vanish at $\mu=0$ for all temperatures due to the CP symmetry of the
QCD partition function and change sign at the endpoint. The
quark-number density $n_q/T^3 = \chi_1$ is always positive for
nonvanishing chemical potential.  Hence, negative values can only be
found at finite $\mu/T$ for $n\geq 3$. The emergence of the first zero
in the ratios is similarly related to the chiral phase boundary as
previously discussed for the even ratios.  It is instructive to define
the distance $\Delta T = T_n - T_\chi$ of the first zero in
temperature direction of the ratio $R_{n,2}$ to the crossover
temperature $T_\chi$.  In \Fig{fig:oddroots_pb} the distance $\Delta
T$ of four different ratios $R_{n,2}$ is plotted as a function of
$\mu/T$ in no-sea MFA.  All curves in the figure start at $\mu/T
\approx 0$ except the one for the ratio $R_{3,2}$ which is positive at
small chemical potentials. All curves obey a minimum whose location
and depth depend on the model and used approximation.  This is
clearly visible in the next \Fig{fig:oddroots_pbVT} where the vacuum
term has been included. For completeness we have also added some even
ratios in the figure which behave very similar to the odd ratios. Thus
our considerations are not restricted to odd ratios.

For the PQM model [10(a)] the negative regions are closer to the
crossover line than for the QM model [10(b)]. All curves
converge to the CEP.  All distances are negative for $n \geq 5$
meaning that the first zero in the ratio $R_{n,2}$ is pushed in the
hadronic phase. The ratio $R_{3,2}$ stays close to the
crossover line and becomes negative only in the vicinity of the CEP,
in particular when the vacuum term is included,
cf.~\Fig{fig:oddroots_pbVT}. For this case the minimum is deeper by
almost a factor of 10 due to the smoothening of the crossover driven
by fluctuations. On the other hand, the Polyakov loop sharpens the
transitions and consequently the minimum is not as deep (a factor 4
smaller) as in the QM model.

An interesting observation is the almost linear behavior of $\Delta T
$ for intermediate $\mu/T$ ratios in both models with and without the
vacuum term and for all ratios. The linear extrapolation of $\Delta T$
from intermediate $\mu/T$ ratios to larger ratios where $\Delta T$
vanishes might serve as an estimator for the proximity of the thermal
freeze-out to the crossover line and the existence of a possible
endpoint can be ruled out for smaller $\mu_c/T_c$ ratios. Of course,
due to some nonlinearities near the CEP this is only a lower bound for
$\mu/T$. The deviations from the linear behavior are yet smaller in
the more realistic PQM model in contrast to the QM model
(cf.~\Fig{fig:oddroots_pbVT}).  This estimate could be strengthened by
considering only the difference of the subsequent roots in the odd
ratios which is independent of the chiral crossover temperature. In
\Fig{fig:oddrootsroots} the relative temperature distance $\Delta \tau
\equiv T_{n+2}- T_n$ for several odd integers is shown as a function
of $\mu/T$ without the vacuum term.  The curves exhibit a similar
behavior as in \Fig{fig:oddroots_pb} except for the ordering of the
curves with respect to $n$. With increasing $n$ the distance between
subsequent ratios decreases signaling a possible convergence of the
negative regions in the phase diagram. An estimation of the lower
bound of the CEP with a linear fit between $\mu/T=0.4 \ldots 0.7$
$(=1.0 \ldots 1.5)$ for the PQM (QM) model in no-sea approximation
yields already a critical $\mu_c/T_c \approx 0.9$ $(2.0)$ for the
lowest ratios ($n=3$) which is quite close to the actual values
$\mu_c/T_c \approx 1.0$ $(2.4)$. For higher $n$ these estimates become
even better. Translating these numbers to the critical temperature
yields $T_c \approx 175 \MeV$ $(100 \MeV)$ already for the lowest
ratio $R_{3,2}$.

\section{Summary}
\label{sec:summary}

In this work we addressed the role of fluctuations in three-flavor
chiral (Polyakov-)quark-meson models and discussed their
phenomenological consequences. The results obtained in no-sea
mean-field approximations where an ultraviolet divergent part of the
fermion-loop contribution to the grand potential is neglected were
compared with findings of the corresponding cutoff-independent
renormalized models. The influence of the vacuum term on the critical
endpoint in the phase diagram, the size of the critical region around
the endpoint and its criticality with respect to its effect on the
universality in general were elucidated in detail.  Because of the
path dependence of the critical exponent towards the critical
endpoint, the critical region which we define as a projection of
constant normalized quark-number susceptibilities on the
$(T,\mu)$ plane is elongated in the direction of the crossover line.  In
the renormalized models a broadening perpendicular to the crossover
line and simultaneously a shrinking in the direction of the chemical
potential axis were found. The broadening is reduced if the
Polyakov loop is taken into account. The Polyakov loop also shifts the
crossover at vanishing chemical potential and consequently the
critical endpoint to higher temperatures.  On the other hand, in the
renormalized models the location of the endpoint is pushed away from
the temperature- towards the chemical potential axis which is a
genuine effect of fluctuations. This shift can be compensated by
reducing the experimentally insecure sigma meson mass in these models
accordingly.  In addition, the matter backreaction to the pure
Yang-Mills sector in the Polyakov-quark-meson models which represents
a further step beyond the mean-field approximation has also been
implemented and the impact on the location of the endpoint and its
critical region was analyzed. With the matter back-coupling
improvement the chiral light phase transition and the peak in the
temperature derivative of the Polyakov loops coincide over a larger
region in the phase diagram when we synchronize both transitions at
vanishing chemical potential by fixing the Polyakov-loop parameter
$T_0$.  This coincidence increases even further when more
fluctuations are taken into account as seen in recent functional
renormalization group and Dyson-Schwinger calculations.

The scaling properties of the quark-number susceptibility in the
vicinity of the endpoint are not modified by the fermion-loop vacuum
fluctuations and a crossover phenomenon between two different scaling
regimes was clearly seen. These regimes also determine the explicit shape of
the critical region in the phase diagram which is not universal
property but depends on the used thermodynamic variables and
underlying equation of state.

As argued in the introduction, higher-order cumulants or generalized
susceptibilities are more sensitive on the diverging correlation
length and might be an appropriate quantity for the experimental
search of an endpoint in the phase diagram. They are given as
corresponding chemical potential derivatives of the partition function
and are thus related to the Taylor expansion coefficients of the
pressure series.

By applying a novel numerical derivative technique, we could calculate
higher moments to very high orders at finite chemical potential. In
contrast to the HRG model which does not have a phase transition with
singularities, the kurtosis stays positive. In our models and
independent of the used approximation the higher-order moments differ
significantly from the HRG model result along the freeze-out curve due
to the existence of an endpoint. As a consequence the ratios between
various moments develop a negative region in the phase diagram close
to the chiral transition line. Fluctuations smoothen the transitions
and hence the ratios and the negative regions are broader and are
shifted towards the hadronic phase in the renormalized models. All
regions converge at the endpoint.

In order to quantify this general trend we have introduced two new
quantities: the distance of the first zero in temperature direction of
various ratios to the crossover temperature and the relative distance
between subsequent roots in the ratios which does not require the
knowledge of the insecure chiral crossover temperature. Both
quantities have a minimum as a function of the $\mu/T$ ratio.  By
using a linear extrapolation for intermediate $\mu/T$ ratios as an
estimator we could rule out the existence of an endpoint for smaller
$\mu/T$ ratios. This estimator might be useful for lattice simulations
where only the negative regions for the first few ratios for small
densities might be available.

An interesting difference to a corresponding two-flavor PQM
calculation concerns the ratio $R_{4,2}$ at vanishing chemical
potential: without the vacuum term the two-flavor kurtosis develops a
peak at $\mu=0$ near the crossover. The vacuum term smoothens the
transition and the peak disappears finally. In the three-flavor
calculation the peak is already suppressed in the no-sea approximation
and hence is not driven by fluctuations.

It would be interesting to investigate higher moments of the so far
neglected electric charge or strangeness fluctuations in particular
beyond the mean-field level with the functional renormalization
group. Work in this direction is ongoing.

\subsection*{Acknowledgments}

We are grateful to Christian Fischer, Kenji Fukushima, Frithjof
Karsch, Volker Koch and Krzysztof Redlich for valuable discussions.
This work was supported by the BMBF Grant No. 06BI9001, the Helmholtz
International Center for FAIR within the LOEWE program of the State of
Hesse and the Helmholtz Young Investigator Group No. VH-NG-332.

\bibliography{../../literature/qcd}

%

\end{document}